\documentstyle[aasms4]{article} 
\newlength{\headroom}
\newlength{\psfigskip}
\begin {document}

\def\3he{$^3$He}
\def\4he{$^4$He}
\def\6li{$^6$Li}
\def\7li{$^7$Li}
\def\he3{$^3$He}
\def\eg{{\it e.g.}}
\def\ie{{\it i.e.}}
\def\etal{{\it et al.\ }}
\def\hii{H\thinspace{$\scriptstyle{\rm II}$}~}
\def\popii{Pop\thinspace{$\scriptstyle{\rm II}$}~}
\def\popi{Pop\thinspace{$\scriptstyle{\rm I}$}~}
\def\la{\mathrel{\mathpalette\fun <}}
\def\ga{\mathrel{\mathpalette\fun >}}
\def\fun#1#2{\lower3.6pt\vbox{\baselineskip0pt\lineskip.9pt
  \ialign{$\mathsurround=0pt#1\hfil##\hfil$\crcr#2\crcr\sim\crcr}}}

\title{Halo Star Lithium Depletion}

\author{M.H. Pinsonneault \altaffilmark{1}, 
T.P. Walker \altaffilmark{1,2},  
G. Steigman \altaffilmark{1,2}, 
and Vijay K. Narayanan \altaffilmark{1}}

\authoraddr {The Ohio State University, Columbus, OH 43210}
\altaffiltext {1} {Department of Astronomy, The Ohio State Univerity}
\altaffiltext {2} {Department of Physics, The Ohio State University}

\date{????? 199?}

%

\begin{abstract}

The depletion of lithium during the pre-main sequence and 
main sequence phases of stellar evolution plays a crucial role in the 
comparison of the predictions of big bang nucleosynthesis with the  
abundances observed in halo stars.  Previous work has indicated a wide range 
of possible depletion factors, ranging from minimal in standard 
(non-rotating) stellar models 
to as much as an order of magnitude in models which include rotational 
mixing.  
Recent progress in the study of the angular momentum 
evolution of low mass stars (Krishnamurthi \etal 1997a) permits the 
construction of theoretical models 
 which reproduce the angular momentum evolution of low 
mass open cluster stars.  The distribution of initial angular momenta can 
be inferred from stellar rotation data in young open clusters.  In this paper 
we report on the application of these models to the study of lithium 
depletion in main sequence halo stars.

A range of initial angular momenta produces a range of lithium depletion 
factors on the main sequence.  Using the distribution of initial conditions 
inferred from young open clusters leads to a well-defined halo lithium 
plateau with modest
scatter and a small population of outliers.  The mass dependent angular 
momentum loss law inferred from open cluster studies produces a nearly flat 
plateau, unlike previous models which exhibited a downwards curvature for 
hotter temperatures in the \7li - T$_{\rm eff}$ plane.    The 
overall depletion factor for the plateau stars is sensitive primarily to the 
solar initial angular momentum used in the calibration for the mixing 
diffusion coefficients.

The \6li/\7li 
depletion ratio is also examined.
 We find that the dispersion in the plateau and the \6li/\7li depletion ratio 
scale with the absolute \7li depletion in the plateau and we use 
observational 
data  to set bounds on the \7li depletion in  main sequence halo stars.
A maximum of 0.4 dex  depletion is set by the observed dispersion 
and \6li/\7li depletion ratio and a minimum of 0.2 dex  depletion 
is required by both the presence of highly overdepleted halo stars and 
consistency with the solar and open cluster \7li data.  The cosmological 
implications of these bounds on the primordial abundance of \7li are discussed.

\end{abstract}

\section{Introduction}

The status of big bang nucleosynthesis (BBN) as a cornerstone of the hot 
big bang cosmology rests on the agreement between the theoretical predictions 
and the primordial abundances of the light elements deuterium (D), helium-3 
(\3he), helium-4 (\4he), and lithium-7 (\7li) inferred from observational 
data (\cite{YTSSO}; \cite{WSSOK}; \cite{CSTI}).  Comparisons of this type 
often rely on models of 
galactic chemical and/or stellar evolution in order to associate the 
observed abundances of these light elements at the present epoch with their 
predicted primordial values.  Recently the confrontation between prediction 
and observation has come under stricter scrutiny as it appeared that the 
primordial abundance of deuterium inferred from ISM/Solar data was smaller 
than its predicted abundance - the value of which follows from requiring that 
the standard big bang model predictions for \4he are in good agreement with 
the abundance inferred from observations of metal-poor extragalactic \hii 
regions (\cite{Hata95}; \cite{CSTII}; \cite{Hata97}).  Very recent 
measurements of the deuterium abundance along the lines-of-sight to 
high-redshift QSOs(\cite{BT97}), along with a reanalysis of the \4he data 
in light of new observations(\cite{OSS96}), increase 
the tension between prediction and observation.

The role of \7li in these comparisons has been minor due to the large 
uncertainty in the estimate of the amount of lithium destruction during the
lifetimes of \popii 
(\ie, metal poor) halo stars which could accomodate primordial lithium 
abundances ranging from the observed plateau value\footnote{The lithium 
abundances of \popii stars with $T_{\rm eff} > 5800$K and [Fe/H] $< -1.3$ is 
nearly independent of metallicity ($[Li] = 2.25\pm 0.1$, where $[X] = 12 + 
\log y_X$ with $y_X$ the number ratio of X to hydrogen) and, hence, is 
referred to as a ``plateau".} (no depletion) up to a factor of ten larger.  
This range in possible depletion factors is equally 
compatible with primordial lithium abundances which correspond to either 
``low deuterium'' (which favors a primordial lithium abundance a factor of 
3 higher than the plateau value), or the observed \4he (which favors the 
plateau value).

The arguments in favor of minimal lithium depletion are the flatness
of the \popii lithium abundance plateau (at low metallicity and high 
temperature) with respect to both metallicity and temperature and the low 
dispersion in the lithium abundance at fixed T$_{\rm eff}$.   This was 
generally consistent with ``standard'' stellar models 
(\ie, models without rotation which burn lithium via convection during 
pre-main sequence evolution) which 
predicted little depletion (\cite{DDKKR}).  There 
are however some specific areas of disagreement in the comparison of the halo 
star data with standard models.  
The observed dispersion may be greater than that predicted 
(\cite{DPD93}; \cite{Thorburn94}; \cite{Ryan96}) or not (\cite{MPB95};
\cite{Spite96}).  There is a small population of highly overdepleted stars 
which appear normal except for \7li (\cite{Thorburn94}; \cite{NRBD}), and 
the trends with metal abundance appear to conflict with expectations from the 
models (Thorburn 1994, \cite{CD94}).  However, the overall 
agreement between standard stellar evolution theory and observations in halo 
stars is good enough that nonstandard models would probably not be invoked to 
explain this data set. Therefore many investigators have adopted the 
reasonable assumption that the observed \popii lithium abundances are close to 
the primordial value.  

The overall properties of \7li depletion in standard models have been
extensively studied (for a review see \cite{MP97}) and 
there are some model independent
 predictions:  There is partial \7li depletion in the pre-main 
sequence (pre-MS), little or no main sequence (MS) depletion for stars hotter 
than about 5500 K, and there should be little or no dispersion in the \7li 
abundance at a given T$_{\rm eff}$ in clusters.  Unfortunately, the 
observational data obtained from halo stars does not serve as a good 
diagnostic for these models since the initial abundance 
is not known and we have no information on the history of the \7li abundance. 
Instead, one can look to  the Sun and open clusters,  systems with a nearly 
uniform initial lithium abundance ([Li] = 3.2 to 3.4) where detailed 
abundance data is available as a function of mass, age, and composition.  
These \popi data provide stringent tests of theoretical models
and it has been known for quite some time 
(\eg, see \cite{WS65} and \cite{Z72}) that standard models 
fail these tests.   The disagreement between the data and standard
models has increased as both the observational data and the 
standard theoretical models have improved. The open cluster data provide clear 
evidence that the \7li abundance decreases with increasing age on the MS, 
contrary to the standard model prediction that the convective zones 
of these hot ($\geq$ 5500K) stars are not deep enough to destroy \7li 
on the MS.  In addition, there is strong evidence for a 
dispersion in abundance at fixed T$_{\rm eff}$ and unexpected mass-dependent 
effects, such as the strong depletion seen in mid-F stars relative to both 
hotter and cooler stars (the ``Li dip'' of \cite{BT86} - for 
extensive reviews of this issue, see \cite{PH88}, 
\cite{MC91}, \cite{S95}, \cite{Bal95}, and 
\cite{MP97}).  For standard models, these features 
cannot be explained by variations in the input physics (\eg, opacities); 
rather, they indicate the operation of physical processes not
usually included in standard models. 
By extension, these processes could also be 
operating in the \popii stars.

Another potentially useful diagnostic of lithium depletion can be found in 
globular clusters.  Clusters provide samples which are homogeneous in age 
and composition, so one would expect a smaller dispersion in a cluster 
sample than in a field star sample.  This is certainly the case for \popi
stars:  \popi field stars do show a larger
dispersion than open cluster stars (\cite{LHE91}; \cite{FMS96}).  However,
recent observations of Li in globular cluster subgiant and turnoff stars
show the opposite trend and
create some serious difficulties for the minimal \7li depletion scenario for
metal poor stars (\cite{DBK95}, \cite{BDSK97}; \cite{TDRRO97}).  
There are clear star-to-star differences in M92;  out of 3 stars
observed with very similar color, 2 have lithium abundances well below the 
plateau value.  In NGC 6397, 20 stars near the turnoff were observed.  For 
7 with identical B-V color, there is a scatter of a factor of 2-3 in lithium
abundance.  Therefore a standard model treatment of the halo field stars must 
explain why the lithium abundances in globular clusters, but not in field 
stars, are anomalous.

The disagreement between the \popi lithium abundances and the predictions 
of standard stellar models has stimulated investigation of nonstandard stellar 
models.  The most prominent explanations are mixing (either from rotation or 
waves), microscopic diffusion, and mass loss (\cite{MP97}; \cite{MC91}).  For 
\popi G and K stars, microscopic diffusion and mass loss are not  likely 
explanations (\cite{MC91}, \cite{SF92}), which 
leaves mixing as a logical candidate.  Mixing induced by rotation can explain 
many of the overall properties of the \popi data.  Rotational mixing can, on 
relatively long timescales, mix material from the base of the convective zone 
to interior regions of the star where lithium can be burned; both the rotation 
rate and the \7li depletion rate decrease with increased age; and a 
distribution of initial rotation rates can produce a distribution of \7li 
depletions and thus a dispersion in abundance at fixed T$_{\rm eff}$.  

Although models with rotational mixing have the qualitative properties
needed to explain the Pop I and Pop II data, two classes of objections have
been raised: 1)  the uniqueness of the solutions and adequacy of the physical
model; and 2) discrepancies in the quantitative comparison of observation and 
theory.
Rotational models require an understanding of the angular momentum
as a function of radius and as a function of time.  There
have been persistent difficulties in reproducing the surface rotation rates as
a function of mass and time in open clusters (\cite{CDP95a},b; 
\cite{KMC95}; \cite{Bou95}).
In addition, models with internal angular momentum transport from hydrodynamic
mechanisms, such as the ones used in this paper, predict more rapid core
rotation in the Sun than is compatible with helioseismology data
(\cite{CDP95a}, Krishnamurthi et al. 1997a (KPBS), \cite{TST95}).  Either of 
these difficulties could potentially affect the degree of mixing in the models.

Models with rotational mixing predict that a range of initial angular 
momenta will generate a range of lithium depletion rates and therefore a 
dispersion
in abundance among stars of the same mass, composition, and age.  However,
the difficulty in reproducing the angular momentum evolution of low mass
stars has made quantifying the predicted dispersion difficult, and therefore
only qualitative estimates have been made (\eg, see \cite{PKD}(PKD),
\cite{CD94}, \cite{CDP95b}, and \cite{CVZ92}).  There have also been 
mismatches between the
mean trend inferred from the data and from all classes of theoretical models.
Chaboyer \& Demarque (1994), for example, concluded that the observed
mean trend of lithium abundance with T$_{\rm eff}$ in metal-poor halo stars 
was not reproduced in {\it any} class of models, be they standard, rotational,
or with diffusion.  There are also cases,
such as the cool stars in young open clusters, where the observed 
dispersion is large but the rotational models predict little, if any,
dispersion.

The case for or against standard model lithium depletion is not as clear
cut when we consider halo stars.  It is argued that models with significant 
rotationally induced depletion could not produce a flat plateau with limited 
dispersion about the mean plateau abundance (\eg, \cite{BM97}).  Furthermore, 
it was expected that such models would destroy far too much \6li 
(\cite{SFOSW}) in conflict with the observation 
of \6li in HD 84937 (see section 4.2).  On the other hand, standard 
(convective burning) models show lithium depletion trends contrary to the 
\popi data.  At minimum the mechanisms responsible for the \popi \7li pattern 
need to be identified and shown to not affect \popii stars.  It is 
 difficult to construct a model where the nonstandard effects 
completely cancel for \popii stars while still retaining consistency with
the \popi data.  Models with microscopic diffusion predict modest \7li
depletion in plateau halo stars.  In general these same models predict that 
the timescale for changes in the surface \7li abundance decreases with 
increased mass; this produces a downward curvature in the 
\7li-T$_{\rm eff}$ relationship at high temperatures which is not observed.  
Mass loss can counteract this trend (\cite{Sw95}; \cite{VC95}).  In either
case, the net 
effect is modest depletion at the 0.2 dex level.  Rotational mixing can also 
suppress diffusion (Chaboyer \& Demarque 1994, \cite{VC95}).   The 
cancellation of different effects still requires at least some mean \7li 
depletion in the halo stars.  In a recent preprint \cite{VC98} note that
although the surface \7li abundance varies strongly with mass in diffusive 
models, the peak subsurface abundance does not (note that the height of the 
peak is
not preserved in models which include mixing).  They then use the height of 
that peak to constrain the absolute \7li abundance, arguing that the 
appropriate mass loss strips each star down to the region that contains this 
peak \7li abundance.    The survival 
of \6li is also problematic in unmixed models with sufficient mass loss to 
expose the peak abundance; in a model of ours which can be compared to 
Figure 1 in \cite{VC98}, \6li drops to half its surface value in the outer 
0.01 solar masses, well below the comparable mass content in the \7li 
preserving region of 0.02 solar masses and also the peak in the \cite{VC98} 
model.  Further, it is disturbing to use a class of models which do not 
include a fundamental characteristic of stars, namely, rotation, when that 
property has been shown to be capable of affecting the issue being studied.

In our view the most serious source of uncertainty in models with rotational 
mixing has been the understanding of the angular momentum evolution of low 
mass stars.  We will show that many of the difficulties in reconciling 
observation and theory are resolved when models that are consistent with the 
rotation data in open clusters are used.
In order to make definitive predictions of the amount of 
rotational mixing, it is necessary to follow the stellar angular momentum 
histories which requires knowledge of the initial distribution of rotation 
velocities along with  
an angular momentum loss law.  There has been significant recent progress 
in constructing theoretical models consistent with the rotation data in young 
open clusters (KPBS; \cite{CL94}; \cite{KMC95}; 
\cite{BFA97}; \cite{A97}).  With the latest generation of 
models, we can both infer the distribution of 
initial conditions and place strong constraints on the surface rotation as a 
function of time.  This enables us to 
quantify the expected dispersion in lithium abundance and greatly reduces the 
sensitivity of the model predictions to uncertainties in the input physics.  
Those models which accurately reproduce the lithium abundances observed in 
open clusters can then be used to predict lithium depletion in halo stars.  
The general trend is that these ``open cluster normalized'' rotation models 
for halo stars predict more lithium depletion than do the standard models but 
less depletion than that predicted by earlier studies of rotational 
mixing with a less sophisticated treatment of angular momentum 
evolution.

 Our goal in this paper is to constrain the amount of lithium depletion in 
\popii stars using several observables: (1) an estimate of the halo star 
initial rotation rate distribution as derived from open clusters, (2) the 
 absence of large dispersion in the observed lithium abundances of the \popii 
``plateau'' stars, (3) the \6li abundance and/or the \6li/\7li abundance ratio 
in HD 84937.  We argue that, 
individually and in combination, these observables point to \popii halo star 
lithium depletion of at least 0.2 dex (following from the open cluster data) 
but no more than 0.4 dex (a consequence of the narrowness of the plateau and 
of the \6li considerations).  Based on the observational data we can then 
infer the primordial lithium abundance and compare and contrast it with that 
predicted by standard BBN for consistency with the inferred primordial 
abundances of D and/or \4he.

\section {Method and Comparison with Previous Models}

\subsection { Angular Momentum Evolution and Rotational Mixing}

In standard stellar models, lithium depletion is a strong function 
of mass and composition; it also depends on the input physics, particularly
the opacities and model atmospheres used to specify the surface boundary 
condition.
 In our models the standard model physics is the same as described in 
KPBS (\cite{KPBS}): namely, interior opacities from OPAL (\cite{RI92}), 
low-T opacities and model
atmospheres at solar abundance from \cite{Ku91a},\cite{Ku91b}); nuclear 
reaction rates from \cite{BPW95}, including the 
\cite{CF88} \7li(p, $\alpha$)
cross-section; Yale EOS; a solar calibrated mixing length, and $Y=0.235$.
Chaboyer \& Demarque (1994) noted that unusual Li-T$_{\rm eff}$ trends for 
cool metal-poor stars occurred in models using the Kurucz atmospheres and 
could be lessened by using a grey atmosphere.  We therefore ran halo star 
models with grey atmospheres and a mixing length of 1.25 as obtained for 
solar models constructed under the same assumptions.

Stellar models which include rotational mixing require additional input 
physics beyond standard stellar models.  The important new ingredients 
include:  

\begin{enumerate}

\item  a distribution of initial angular momenta 

\item a prescription for angular momentum loss 

\item a prescription for the internal transport of angular momentum and the 
associated mixing in radiative regions

\item the impact of rotation on the structure of the model.

\end{enumerate}
There have been important changes in the 
treatment of the first three ingredients since the study of rotational mixing 
in halo dwarfs by \cite{PDD}; in this 
section we discuss
the ingredients of the current set of models and compare them with 
previous work.  The treatment of angular momentum evolution in this 
set of models is the same as KPBS.  The structural
effects of rotation have been computed using the method of 
\cite{KT70}, and are small for low mass stars which 
experience angular momentum loss.

\subsubsection {Initial Angular Momentum}

The studies of lithium depletion in the presence of rotationally induced
mixing by \cite{PKSD} (the Sun), PKD (open cluster stars), and 
PDD (halo stars) all used a range of initial 
rotation rates in the pre-MS from the measured rotation velocities of young 
T Tauri stars (10-60 km/s) at a typical reference age of 1 Myr.  No 
interaction between accretion disks and the protostar was included; the entire 
range of initial conditions was generated from the range in the initial 
rotation velocity.  This caused some difficulty in reproducing the observed
range in rotation (a factor of 20 in young open clusters), especially since 
the higher angular momentum loss rate in rapid rotators acted to reduce the
range in initial rotation rates.  This approach was also used by 
Chaboyer \etal (1995a,b) for the Sun and open cluster stars respectively and 
by Chaboyer \& Demarque (1994) for halo stars, although these papers used a 
different angular momentum loss law than the earlier work.

As protostars evolve their moment of inertia decreases; this would produce an 
increase in rotation rate with decreased luminosity.
In contrast, the rotation periods of classical T Tauri stars appear to be 
nearly uniform - 
they do not scale with luminosity as they would if the stars were conserving 
angular momentum as they contracted towards the main sequence (\cite{B93}, 
\cite{E93}).  Weak-lined T Tauri stars, which are thought to be protostars
without significant accretion disks, have much shorter rotation periods.  
This behavior would be expected if there 
were an accretion disk regulating the rotation of the central object and
spinup only occurred when the disk is no longer magnetically coupled to the
protostar (\cite{K91}; \cite{CCQ95}; \cite{KMC95}).  In this revised picture 
there is a constant surface rotation rate while there is a sufficiently 
massive disk around the central object; stars which detach from their disks 
earlier experience a larger change in angular velocity as they contract to 
the main sequence than those which detach from their disk later (and are only 
free to begin to spin up when their moment of inertia is smaller).  This 
disk-locking hypothesis leads to a larger range in rotation rates for 
models of young open cluster stars, in better agreement with the data, 
and explains the very slow rotation of the majority of young stars 
(\cite{BFA97}; \cite{CCQ95}; \cite{KMC95}; KPBS).
These differences in the initial conditions lead to interesting 
consequences for the lithium depletion pattern.  Although the initial rotation 
periods are comparable to those used in the earlier studies, the 
inclusion of star-disk coupling implies that the MS angular 
momentum is much lower for models with long disk lifetimes than it would be if 
the disks were absent.  Disk lifetimes of 3 Myr are needed to match the 
rotation rates of the slow rotators in young clusters.  The mean depletion for
the majority of stars is 
therefore significantly lower than in previous studies because most young 
stars are slow rotators and the degree of mixing 
decreases for models with lower MS angular momentum.

\subsubsection {Angular momentum loss.}

PKSD, PKD, and PDD all used a loss law of the  form ${dJ}/{dt} 
\propto \omega^3$, where $\omega$ is the angular velocity (\cite{K88}).  
This implies greater angular momentum loss for rapid rotators than for slow 
rotators.  However, models of this type do not reproduce the rapid rotators seen in 
young open clusters; the spin down of high angular momentum objects is too 
severe (PKD).  On both theoretical and observational grounds, a more realistic 
functional form for the loss law is

\begin {equation}
\frac{dJ}{dt} \propto \omega^3\ \ \ \ \ \ (\omega < \omega_{\rm crit}),
\end {equation}
and
\begin {equation}
\frac{dJ}{dt} \propto \omega \omega_{\rm crit}^2 \ \ \ \ \ \ 
(\omega > \omega_{\rm crit})
\end {equation}
(\cite{MB91}; \cite{CCQ95}; \cite{M84}; \cite{BS96}).
Here $\omega_{\rm crit}$ is the angular velocity at which the angular 
momentum loss rate saturates; $\omega_{\rm crit}$ can be estimated from 
several observable quantities, thought to be correlated with surface 
magnetic field strength, which saturate at 5-20 times the solar rotation 
period (see \cite{PS96}, \cite{Kr97b} for reviews and discussions of recent 
work).  A loss law which saturates at high rotation rates allows rapid 
rotation to survive into the early main sequence phase; a loss law of this 
form was adopted by Chaboyer \etal (1995a,b) and Chaboyer \& Demarque (1994).

The observational data indicate that the duration of the rapid rotator 
phase is a function of mass in the sense that rapid rotation survives longer 
in lower mass stars.  There is also indirect observational evidence 
for a mass-dependent saturation threshold (\cite{PS96}; 
\cite{Kr97b}).  Models where the saturation threshold scales with 
the convective overturn time scale can reproduce the observed mass dependent 
spin down pattern.  We adopt the mass-dependent saturation threshold of 
KPBS,

\begin {equation}
\omega_{\rm crit} = \omega_{\rm crit}(\odot) \frac {\tau_{\rm conv}(\odot)} 
{\tau_{\rm conv} (*)},
\end {equation}
where the convective overturn time scale $\tau_{\rm conv}$ as a function of 
ZAMS T$_{\rm eff}$ was taken from the 200 Myr isochrones of Kim \& Demarque 
(1996; KD).  \cite{KD96} only considered models at solar abundance, and we 
would need to recalibrate the KPBS models if we used a time and abundance 
dependent overturn timescale for the models.  A comparison of the convection 
zone depth (as a function of T$_{\rm eff}$, Z, and age) in the tables of
PKD and PDD indicates that at young ages the convection
zone depth at a given T$_{\rm eff}$, and therefore the overturn time scale, 
depends only weakly on metal abundance.  The value of $\omega_{\rm crit}$ is 
important primarily in the early main sequence.  We therefore evolved 
standard halo star models to an age of 200 Myr and used the same 
$\tau_{\rm conv}$ as a function of T$_{\rm eff}$ for the
metal-poor models as reported by KD for the solar abundance
models.  We will consider models with a mass, composition, and time dependent
overturn time scale in a future paper in preparation (\cite{N98}).
A loss law with a mass-dependent saturation threshold acts preferentially to 
suppress rapid rotation in the more massive stars, unlike the constant
saturation threshold adopted by Chaboyer \& Demarque (1994).  This reduces 
both the absolute depletion and the 
dispersion of lithium abundances in hotter halo plateau stars relative to
cooler plateau stars.

\subsubsection {Angular momentum transport.}

The treatment of angular momentum transport is the same as in KPBS; a detailed 
review of the time scale estimates can be found
in Pinsonneault (1997).  We consider internal angular momentum transport by
hydrodynamic mechanisms alone, and do not include potentially important
mechanisms such as gravity waves (\cite{KQ97}; \cite{ZTM97})
and magnetic fields (\cite{CM92}, \cite{CM93}; \cite{BCM98}).  In 
previous work the degree of mixing was found to be 
insensitive to the assumptions about internal angular momentum transport 
(PKD; PDD; \cite{CDP95a},b).  \cite{CVZ92} and \cite{R96} obtained similar 
depletion patterns in models with solid body
rotation, and \cite{BCM98} reported similar results
in models with magnetic fields and hydrodynamic mechanisms.  This can be
understood because the diffusion coefficients for angular momentum transport
are actually largely determined by the balance between the surface boundary 
condition (the applied torque) and the flux of angular momentum from below a
given shell.  Theory gives an estimate of the ratio of the diffusion 
coefficients for mixing to those for angular momentum, which is related to 
the existence of anisotropic turbulence in stellar interiors (\cite{CZ92}).  
The diffusion coefficients for mixing are calibrated on the
Sun, but our lack of information on the rotational history of the Sun requires
an exploration of models with varying solar initial conditions.

\section{Results}

\subsection{Initial Conditions and Angular Momentum Evolution}

We have taken the Pleiades $v\sin i$ data used in KPBS and applied a
statistical correction of 4/$\pi$ to all of the data points to correct
for inclination angle effects.  A grid of solar composition models with a 
range of accretion disk lifetimes was constructed and evolved to the age of 
the Pleiades.  We interpolated in the grid to map the distribution of 
measured rotation velocities onto a distribution of accretion disk lifetimes.
A histogram of the number of stars as a function of disk lifetimes is 
presented in Figure 1a.  This distribution of initial conditions was used to
infer the distribution of lithium depletion factors for different solar disk
lifetimes in all of the following discussion.  We then generated a grid of
halo star models using the same input physics; we compare the rotation 
velocities of Pleiades stars to those for halo composition stars with the
same disk lifetimes in Figure 1b.  The distribution of initial conditions will
be important for the remainder of our results, so we begin by discussing the
properties of the Pleiades sample first.  We then compare the angular momentum
properties of our halo and solar abundance models.

\subsubsection{Distribution of Initial Conditions}

The measured rotation rates in the Pleiades are subject to ambiguity because
we measure $v\sin i$ rather than the true rotation rate; the true rotation
velocities will therefore be higher on average than the $v\sin i$ values.
In addition, there are a significant number of upper limits in the data set;
this technique would therefore miss any population of very slow rotators.
Such stars, if present, would have less lithium depletion than our technique
would permit, and they would therefore influence the predicted range in
lithium depletion factors.  Finally, there is the possibility that the
distribution of initial conditions could depend on environment; \ie, the
assumption that halo stars had the same distribution of disk lifetimes as
the Pleiades, or that the Pleiades is the same as other open clusters,
may be incorrect.

A distribution of $v\sin i$ values can in principle be inverted to 
infer the true distribution of velocities (\cite{CM50}), but 
the technique requires a large sample and we would lose the information on 
the mass dependence of rotation which is important for our purposes.
Although $\sin i$ can be small, it will be close to 1 for the large majority of
stars; only 1/7 of the sample will have $\sin i<0.5$ for a random 
distribution of inclination angles.  Furthermore, the young open clusters
have a small population of rapid rotators and a large population of slow
rotators; 13 out of the 91 stars in the Pleiades sample have $v\sin i>40$
 km/s, and 7 have $v\sin i$ between 20 and 40 km/s.  Therefore, only a small 
fraction of the low $v\sin i$ systems are intrinsically rapid rotators seen
pole-on.  For these reasons we have therefore chosen to apply an average 
statistical correction to the data, and do not believe that the shape of the
distribution in Figure 1 will be greatly altered by inclination angle effects.
Rotation periods are not affected by $sin i$, and they provide an alternate
means of estimating the distribution of initial conditions.  Although the 
present rotation period samples are much smaller than the $v\sin i$ samples,
the number of stars with measured rotation periods is growing and it may soon
be possible to use them in well-studied systems such as the Pleiades.

The issue of the lower envelope of rotation is a more serious concern; there
is a bias in the $v\sin i$ sample against detecting truly slow rotators.  We
have a series of reasons for believing that the absence of very slow rotators
is real.  First, stellar rotation periods can be measured for stars without
the uncertainty introduced by the inclination angle; much slower rotation rates
can be measured with this technique than with $v\sin i$ data.  Long
rotation periods are not found in the Pleiades (\cite{Kr97b})
and are very rare in the T Tauri population (\cite{E93}, \cite{B93}), 
which suggests that any population of very slow rotators is
small.  Furthermore, even disks that last to the main sequence would only
produce stars with rotation periods of order 10 days, very close to the
rotation period corresponding to the $v\sin i$ limits.  We would therefore
require stars which are locked to their disks with a period much longer
than that seen in T Tauri stars and with a long disk lifetime to produce a
population of ultra-slow rotators large enough to affect our results.
Finally, progressively younger open clusters have a smaller number of upper 
limits than older systems (there are fewer in $\alpha$
Per than in the Pleiades), and the lower envelope of rotation is resolved
in the youngest systems (IC 2391 and IC 2602).  Recent studies of rotation in
the Pleiades are also consistent with this conclusion (\cite{BFA97}).

The universality of the distribution in Figure 1a is another important
question.  KPBS found that the distributions of disk lifetimes needed for the
different young open clusters were consistent with one another, in the sense
that the slow rotator distribution peaked at similar disk lifetimes for
systems of different ages.  This does not
guarantee that the same will hold true for systems that formed under
different conditions, such as halo stars and globular cluster stars.  The 
distribution of initial rotation rates cannot be directly inferred in old
systems because the surface rotation rates of stars converge as they age and
lose angular momentum.  Even in a system as young as the Hyades (600 Myr) the
range in surface rotation rates is too small to permit a reliable estimate of
the initial angular momentum of any given star.  It is,
however, at least a plausible starting point to ask whether the distribution
of lithium abundances in different samples is {\it consistent} with the 
distribution of rotation rates seen in young open clusters such as the 
Pleiades.  One can in principle invert the question and ask what distribution
of disk lifetimes is needed to produce a given distribution of lithium
abundances.  Lithium depletion depends on the rotational history of a given
star, and therefore the distribution of surface lithium abundances in old
stars can be used to obtain information about the distribution of
initial conditions in systems that are too old to provide direct information
on the rotation rates.  We will show that
both the halo star and open cluster lithium depletion patterns are consistent
with lithium depletion from rotational mixing and with being drawn from the 
same set of initial conditions.

\subsubsection{Metallicity Effects}

For each Pleiades star we can infer a disk lifetime; in Figure 1b we compare 
the rotation properties of halo star models with the same age, effective 
temperature, disk lifetime, and rotation physics (angular momentum loss law 
and angular momentum tranport prescription).  We did not run halo star models
with effective temperatures as low as the bulk of the Pleiades sample, since
these stars would not be on the halo lithium plateau; we have therefore only
shown the T$_{\rm eff}$ range where the two samples overlap.  There are some
important features of this comparison which will have implications for the
lithium depletion pattern in open cluster and halo stars.  First, the 
overall level of rotation is lower in the halo stars; second, the range in
rotation rates is smaller.  Both phenomena can be traced to differences in
stellar structure, and the relative pattern is the same as described in
PDD and in Chaboyer \& Demarque (1994).

At a given 
T$_{\rm eff}$ a halo star will have a lower total mass and moment of inertia;
for the same amount of angular momentum loss the halo star will therefore
spin down more rapidly.  Rapid rotators lose more angular momentum than slow
rotators, which causes their surface rotation rates to converge at late 
ages.  In addition, the convergence to similar surface rotation rates occurs 
more quickly in the halo stars because of their
smaller moments of inertia.  The diffusion coefficients for mixing 
depend on both the absolute rotation rates and the internal angular velocity
gradients; both are smaller in halo stars than in solar composition stars of
the same age and T$_{\rm eff}$.  The overall level of lithium depletion will 
therefore be lower in the halo star models.  In addition, a range in 
surface rotation rates will produce a range in the degree of rotational 
mixing, and therefore a dispersion in surface lithium abundances.  The 
smaller range in surface rotation rates for the halo star models will
therefore produce a smaller dispersion in surface lithium abundance at
fixed T$_{\rm eff}$.  Similar initial conditions will therefore produce 
different apparent lithium depletion patterns; this is important because
such differences are clearly seen in the observational data.

\subsection{Monte Carlo Simulations of Open Cluster Lithium Depletion}

Our theoretical models can be used to predict lithium depletion in open 
cluster and halo stars.  As discussed above,  the observed distribution 
of rotation rates in young open clusters (like the Pleiades) is mapped 
onto a distribution of initial rotation rates by use of an accretion 
disk model which, during 
the T Tauri phase of its pre-main sequence evolution, locks the rotation 
period of the star at 10 days.  A range in decoupling 
times for these accretion disks will then generate a range of rotation 
rates for stars on the main sequence as a function of time.  The rotational
properties of the models are calibrated on the rotational properties of
the open cluster stars and solar models are required to reproduce the
solar rotation rate at the age of the Sun.  This step is independent of the 
initial conditions for the Sun because different disk lifetimes for solar 
models converge to identical surface rotation rates at the age of the Sun.

We then need a means of relating the diffusion coefficients for mixing to
those for angular momentum; the usual approach is to require that a solar
model reproduce the solar surface lithium abundance at the age of the Sun.
Because models with different disk lifetimes have different rotation histories,
this step depends critically on the disk lifetime assigned to the Sun.
If the Sun is assumed to have a typical rotation history then solar analogs 
will have
lithium depletion comparable to that in the Sun.  On the other hand, if the 
Sun was an unusually
rapid rotator (\ie, had a shorter disk lifetime than most stars) the majority
of other stars will have
less lithium depletion.  We therefore ran solar calibrated models with the
Sun having disk lifetimes of 0 (anomalous), 0.3, and 1 (``typical'') Myr; 
these cases will be referred to as s0, s0.3, and s1 for the remainder of 
the text.  The open cluster lithium data are fit poorly when the Sun is 
assumed to have a disk lifetime longer than 1 Myr (see below).  All of our 
solar disk lifetimes are shorter than the 3 Myr required for typical stars; 
this implies that the Sun was a faster than average rotator in its youth and 
thus its rotation history was not ``typical'', strictly speaking.  
The solar-calibrated models can then be used to predict the lithium depletion 
factor as a function of time for a given disk lifetime, mass, and composition.  
We then investigated the distribution of lithium depletion factors that would 
be produced by a given distribution of disk lifetimes.  For each star we chose
an initial condition at random from the Pleiades distribution of disk
lifetimes; we show sample Monte Carlo runs for the four open clusters using
two different solar calibrations (s0 and s1) in Figures 2 and 3 
respectively.  For each cluster all stars were assumed to start with the same 
initial lithium abundance: we chose [Li] = 3.4 for the youngest open clusters 
and [Li] = 3.3 for the Sun and for M67.  Standard model predictions for Li
in these clusters are shown as solid curves in Figures 2 and 3.  Note that 
some main sequence lithium depletion is demanded by the data and that the
overall lithium depletion pattern as a function of age is in reasonably good
agreement with the Monte Carlo simulations of the models with rotational 
mixing.  Variations in the initial rotation rates provide a natural 
explanation for dispersion in stellar surface Li/H at fixed T$_{\rm eff}$.  

The predictions of the time dependence of lithium depletion 
due to rotational mixing matches the open cluster data best for the
case where the Sun was an unusually rapid rotator (Figure 2). 
This s0 case corresponds to minimal lithium depletion and predicts modest 
dispersion in the open cluster lithium abundances.  As the Sun is made more 
typical (\ie, as its disk lifetime increases and ZAMS rotation rate drops) 
the overall depletion and dispersion become larger and the overall
agreement with the data worsens.  The intermediate-aged systems are the
most sensitive discriminant; the model predictions are similar in young
systems and the old clusters have less data (which is also less accurate).
  A more detailed analysis of the open cluster 
lithium depletion pattern will be presented in a paper in preparation
(\cite{N98}); for the purposes of this paper, we note
that open cluster models constructed using the same technique that we
have applied to halo stars are in overall agreement with the observed
lithium depletion pattern.

\subsection{Halo Stars}

\subsubsection{Parameter Variations}

We can extend the PNK models, designed for rotational mixing in open clusters, 
to study lithium depletion  in halo stars.  Unlike the open clusters, the
halo stars have a range in metal abundance and age.  We therefore begin with
a discussion of the dependence of the lithium depletion pattern on the
important components of our models: disk lifetimes, solar calibration,
metal abundance, and age.  In Figure 4 we show the effects of changes in
all of these ingredients.
In all panels the solid line is a base case which corresponds to models 
with a 3 Myr disk lifetime, $[Fe/H]=-2.3$, and an age of 12 Gyr.

\noindent 1.  Disk lifetimes.

A range in disk lifetime can produce significant variations in the predicted
lithium depletion, especially for the rapid rotators.  This will manifest
itself as a dispersion in lithium abundance at fixed T$_{\rm eff}$ (top panel
of Figure 4).  However, the large majority of stars in the Pleiades are slow 
rotators with disk lifetimes greater than 1 Myr.  As we will see below, the 
intrinsic range in depletion factors for slow rotators is very small; \ie, 
only the top 20\% of the distribution will have depletion factors greater 
than or equal to the depletion factor of the 1 Myr case see Fig.~1a), while 
the remainder will have depletion factors similar to that for the 3 Myr case.  
The overall pattern that would be expected from the Pleiades distribution of 
initial conditions is therefore a majority of stars with a small intrinsic 
range in abundance with a subpopulation (corresponding to rapid rotators) 
which is overdepleted with respect to the bulk of the stars.

\noindent 2.  Solar Calibration.

The solar calibration sets the overall level of depletion in the models,
and it will therefore have a strong influence on the derived primordial
lithium abundance (the second panel of Figure 4).  Once the solar calibration 
is set by the open cluster data, it should be the same for all stars 
(\cite{CZ92}), and in particular for halo stars.  That is, we 
expect that the best fit to the halo star lithium abundances will correspond 
to the limit in which the Sun is over-depleted in lithium relative to its 
counterparts (the s0 case).

We define the depletion factor $D_7$ as the ratio of the initial (\7li/H) to 
the current surface (\7li/H).  The halo star models of PDD 
predicted a value of $D_7$ close to 10 for plateau stars, 
with little dependence on the model parameters.  This behavior is very
different from that illustrated in Figure 4, and this can be traced to
differences in the initial conditions and angular momentum loss law.  PDD 
began with a modest 
range of initial angular momenta (a factor of 10) and applied a loss
law which caused a rapid convergence in the surface rotation rates.  We 
begin with a wider range of initial angular momenta (disk lifetimes from
0 Myr to 10 Myr will produce a factor of 16 range in angular momentum at
an age of 10 Myr).  The loss law that is required by the rotation data also
permits a wide range of rotation rates to persist for a longer period of time
than it did in the models of PDD.
The net result is a much larger possible range of lithium depletion factors.
The models of PDD, for example, had a total range in log($D_7$) of 0.25 dex at
$0.7 M_{\sun}$ and [Fe/H]=$-2.3$, while the corresponding models in the
current study have a range in depletion factors of 0.50 dex (s0) to 1.30 dex
(s1).  Because the possible range of lithium depletion factors is greatly 
increased, the choice of the initial conditions for the Sun becomes 
correspondingly more important.  In effect, the earlier generation of models
had a small enough effective range in initial conditions that the Sun was
by definition a typical star and the typical depletion factors were therefore
forced to be similar to those for the Sun.  If the Sun is treated as a typical 
star with a disk lifetime of 3 Myr, we recover the mean depletion factor of 
PDD ($\log (D_7)= -1.0$) with a much larger range of depletion factors than 
permitted by the data.

\noindent 3.  Metal Abundance

Models with different metal abundance differ in their overall lithium
depletion factors by a modest amount, less than 0.1 dex change in lithium
depletion per 1.0 dex change in [Fe/H].  The overall trend is in the sense that
the more metal-rich stars deplete more than the more metal-poor stars; this
is to be expected in light of the relative differences between solar
composition and halo star rotational properties illustrated in Figure 1.

This is opposite to the trend found by Thorburn (1994) in the observational 
data, for which stars with lower metal abundance have, on average, slightly 
lower lithium abundances, by 0.1 dex in [Li] for 1.0 dex in [Fe/H].
However, there are some possible complications in this relative comparison.
The slope of the [Li]-[Fe/H] relationship depends upon the assumed
model atmospheres.  Furthermore, the derived value of the slope depends on
the data analysis method which is used and is different for different 
subsets of the data (\cite{Ryan96}).  By contrast, the
[Li]-T$_{\rm eff}$ slope reported by Ryan \etal (1996) is resistant to these 
effects.

Within the framework of the rotational mixing hypothesis, one
would also expect stochastic variations in the disk lifetimes and therefore
an additional random component in the lithium abundances.  Since about 20\%
of the Pleiades stars have modest to rapid rotation, the sample could
(for example) happen to have slighty more rapid rotators at low [Fe/H] and
slightly fewer at relatively high [Fe/H] by chance; this effect is not
included in the Ryan \etal (1996) or Thorburn (1994) analysis.

Another possibility is that the trends in the models and the data are 
correct, and that Galactic production of \7li is responsible for the 
difference 
between the trends.  In all of the models we require some lithium production
during the lifetime of the Galaxy; this would imply that the more metal-rich
stars begin with slightly more lithium, which would counteract the 
[Li]-T$_{\rm eff}$ trend predicted by the models.
The Galactic evolution of lithium remains a mystery, but we can estimate the
effect as follows.  If we scale the solar system lithium abundance
([Li]=3.3) linearly with metallicity the newly produced lithium will
contribute $\Delta$[Li]~=~0.0, 1.0, 2.0 at $[Fe/H]=-3.3, -2.3, -1.3$ 
respectively.
Note that if the post-BBN lithium production scales faster than linearly
with metallicity such halo contamination will be negligible.  We infer a
primordial abundance in the range [Li]$~=2.35-2.75$.  The total abundance
at the low end of the range
(primordial plus production) would therefore be 2.35, 2.37, and 2.51 at
$[Fe/H]=-3.3, -2.3, -1.3$ respectively; by comparison, at the high end of
the range the comparable abundances would be 2.75, 2.76, and 2.82.  This
suggests that galactic production could influence the Li-[Fe/H] slope at 
higher metal abundances but not for abundances below $-2.3$.

As we will show
below, the detection of \6li in HD 84937 requires some \6li and \7li
production at early epochs; this production could have been highly variable
and possibly even largely absent for the most metal poor stars.  We therefore 
cannot exclude an additional
production component between the very metal-poor stars
and the more typical plateau stars, but caution that there may not be a
reliable relationship between production and metal abundance in this case.  
If the level of \6li seen in HD 84937 is typical for stars
at [Fe/H]=$-2.3$, and the production of 
\7li by $\alpha-\alpha$ fusion is comparable to that of \6li, then we can
set an upper limit on the variation in the initial \7li abundance between the
most metal poor stars and the typical plateau stars by computing the \7li
yield required to produce the observed \6li in HD 84937.  The \7li yield
required to explain the observed \6li in HD 84937 is in the range of
3 $\times 10^{-11}$ to 7 $\times 10^{-11}$.  This suggest
that variable production of \7li could have a modest impact on the 
[Li]-[Fe/H] slope even at early epochs. 

In light of the uncertainties discussed above we have adopted an empirical
approach.  The value of the [Li]-[Fe/H] slope which minimizes the dispersion
in the data set is computed, and the observed lithium abundances are then 
corrected for the effects of metallicity.  We have therefore used the 
$[Fe/H]=-2.3$ models for our theoretical 
calculations and corrected the Thorburn (1994) abundances to a uniform
[Fe/H] at this level using the relationship of Thorburn (1994),

\begin{equation}
[Li]_{\rm detrended} = [Li]_{\rm observed} + 0.13([Fe/H]+2.3).
\end{equation}

This raw data is compared with the data corrected to a uniform abundance of
$-2.3$ in Figure 5.  We note that another approach would be to draw the 
distribution of theoretical depletion factors from the distribution of
[Fe/H] in the Thorburn sample, \ie, to Monte Carlo both the initial conditions
and the [Fe/H] abundances.  In numerical tests the correction in equation (4)
reduces the dispersion in the data set by an
amount comparable to the level introduced by the observed distribution
of metal abundances convolved with the theoretical models.

\noindent 4.  Age.

The final ingredient is age; as shown in the bottom panel of Figure 4 even
significant variations in age produce only small changes in the lithium
abundance at fixed T$_{\rm eff}$.  This results from a combination of a low rate
of lithium depletion at old ages and the evolutionary changes in T$_{\rm eff}$ as
a function of time, which move models along the plateau.  We therefore 
neglect the impact of age variations on the dispersion and depletion patterns.

\subsubsection{Monte Carlo Simulations:  Trends with T$_{\rm eff}$ and Dispersion}

Monte Carlo simulations of the lithium abundances of halo stars are compared
with the Thorburn (1994) dataset detrended to a uniform [Fe/H] of $-2.3$ in 
Figure 6.  The 
different models all show a [Li]-T$_{\rm eff}$ trend similar to the data.  
Previous models with rotational mixing had a downward curvature at the hot 
end of the plateau, which is highly suppressed in these models, although
there is
a slight downward curvature in the s1 case.  This difference can be attributed
to improvements in the angular momentum loss law.  The open cluster rotation
data requires stronger angular momentum loss for higher mass stars; this
acts to preferrentially suppress lithium depletion in the higher mass models
at the hot end of the plateau relative to models for cooler lower mass
stars.  

The Pleiades distribution of initial conditions also produces a mean trend with
small internal scatter and a smaller population of overdepleted stars.  As
the solar disk lifetime is increased the dispersion and overall depletion both
increase; we can therefore use the dispersion in the halo plateau to set
constraints on the overall depletion.  The simulations in this section do
not include observational errors, whose impact on dispersion estimates
will be discussed in the next section.  The Thorburn sample also contains a
small population of highly overdepleted stars; their existence is a problem
for standard stellar models and they are usually ignored in discussions of
the dispersion in the halo plateau.  We will not ignore the
outliers; in fact, such objects are a natural consequence of the models with 
rotational mixing combined with the observed distribution of rotation rates
in young clusters.  The maximum depletion
level for the rapid rotators also increases as the absolute depletion
increases.  The existence of highly overdepleted stars can therefore
be used to constrain the minimum level of absolute depletion in the halo
plateau; if the absolute depletion in the halo plateau is too small
then stars far below the plateau should not be present, but they are.

Both the trends in the [Li]-T$_{\rm eff}$ plane and the dispersion in the 
plateau
have been used to argue against significant lithium depletion in halo stars.
The simulations in Figure 6 indicate that the current class of models are
compatible with both, but the agreement between the data and the models
worsens if the absolute depletion is too high.  Similar trends are present
in the open cluster simulations.  {\sl In other words, the open cluster
and halo star data are consistent with being drawn from a ZAMS angular 
momenta distribution similar to that required to explain the distribution
of rotation rates in young open clusters.  Furthermore, both are best 
explained by a solar calibration in which the Sun had an accretion disk
lifetime shorter than the typical value for low mass stars.}

In the next section we quantify the constraints placed on the overall
lithium depletion by the dispersion and the highly overdepleted stars.
We also discuss the relative lithium-6 and lithium-7 depletion factors and
constraints on the overall destruction of lithium-7 from the detection of
lithium-6 in HD 84937.

\section{Constraints on Rotational Mixing from Halo Stars}

\subsection{Dispersion of the Plateau Abundances}

Lithium abundances versus effective temperatures for warm ($T_{\rm eff}\ga 
5800$K), metal-poor ([Fe/H] $\leq -1.3$) stars (as observed by
\cite{Thorburn94}) were shown in Figure 5.  Four independent groups have 
observed \popii plateau stars (\cite{Spites}; \cite{Spite84}; \cite{Rebolo88};
\cite{Thorburn94}).  The differences in the lithium abundances 
derived by different observers from \popii halo stars can be traced to the 
choice of temperature scale and of model 
atmosphere required to extract an abundance from an observed 
absorption line.  A survey of the literature shows that there may be a 
systematic uncertainty in the lithium abundance scale of order $\pm 0.1$ 
dex due to the choice of atmosphere model and of the stellar temperature 
scale.  The  intrinsic star-to-star dispersion is likely to be much 
smaller than this.  There may also be correlations in the data in the 
sense that the lithium abundance increases slightly with effective 
temperature and with metallicity (\cite{Thorburn94}; \cite{Ryan96}) or not
(\cite{MPB95}).  We adopt [Li]$_{\rm PopII} = 2.25\pm 0.10$ 
as the weighted mean plus estimated
{\it systematic} error of all the plateau data.
The raw dispersion about the trend is 0.16 dex (Thorburn 1994).   
When the T$_{\rm eff}$ and [Fe/H] trends are removed, Thorburn estimated
the dispersion as 0.13 dex; by comparison, her estimate of the observational
error is 0.09 dex.  Estimates of the
intrinsic star-to-star dispersion are therefore roughly half the raw
dispersion ( 0.08 dex (\cite{DPD93}),  0.06 dex (Thorburn 1994), 
 $<0.1$ dex (\cite{BM97})).  
  
In our rotationally mixed models, the ``anomalous 
Sun'' case corresponds to lithium depletion of 0.2 dex and a dispersion 
in depletion of 0.09 dex, consistent with estimates of the intrinsic
dispersion.  The intermediate disk case corresponds to lithium depletion 
of 0.4 dex and a dispersion 
in depletion of 0.16 dex which is at the limit of the raw dispersion.
The ``typical Sun'' case corresponds to lithium
depletion of 0.69 dex and a dispersion of 0.25 dex which is considerably 
larger than the raw dispersion.  This initial comparison suggests that an  
upper-bound to lithium depletion in halo stars is 0.4 dex.
If the Sun is anomalous (as favored by the open cluster data) then little 
dispersion in the plateau is predicted, consistent with the data, along 
with a smaller overall depletion of 0.2 dex.  So, two indicators, the open 
cluster data along with the lack of significant dispersion in the halo lithium
abundances, both point to halo star lithium depletion in the range 0.2 - 0.4 
dex.  

However, the calculation of the observational dispersion excludes 6 upper 
limits, and the distribution of theoretical depletion factors
is significantly non-gaussian.  We can use these additional characteristics 
of the plateau data to constrain the
absolute depletion.  Standard models would require that 
the only dispersion in halo star lithium abundances is due to age, metal
abundance, and observational error.  Age is a small effect in all classes of
models, and the effects of metal abundance can be included by detrending the
data.  We would therefore expect the shape of the distribution to be determined
by the observational errors and thus it should be gaussian.

By contrast, the distribution of lithium depletion factors in rotational models
is skewed.  The majority of young cluster stars are slow rotators, so the
majority of the depletion factors are tightly clustered with a very small
intrinsic dispersion; there is a component of stars with moderate rotation
rates which produce a population with larger depletion factors and a small
fraction (2-3\%) of rapid rotators with large depletion factors.  In
addition, the distribution will be smeared out by observational errors.
We ran 1000 Monte Carlo simulations at intervals of 50K across the plateau and
drew theoretical lithium depletion factors from the Pleiades distribution of
initial conditions.  In one set of runs observational errors were neglected;
in a second set of runs we added a random error of 0.09 dex.  The lithium
depletion factors for each T$_{\rm eff}$ were then sorted. 
In Figure 7 we compare the detrended Thorburn data (bottom panel)
with the 2.5\%, median, and 97.5\% bounds of the theoretical depletion 
factors for
our s0 case (top), s0.3 case (second panel) and s1 case (third panel).
The zero point was set by requiring that half the stars in the sample for
all three cases were above the median and half the stars were below it.
Figure 8 presents the same cases for simulations which include observational
errors.  The top and bottom dashed lines can be regarded as effective
$\pm 2\sigma$ bounds.  {\it Note that the observed lithium abundances have 
a similar morphology to the simulations; in particular, they are not 
gaussian distributed about the mean as would be expected if these stars 
had suffered no lithium depletion.}  Furthermore, the intrinsic dispersion 
for the majority of stars is significantly less than the observational 
errors; in other words, the dispersion for the majority of stars is 
dominated by observational errors.  The theoretical models predict that 
there should be a small number of highly overdepleted stars, which are 
indeed observed.

This can be further quantified by comparing the distribution of abundances
about the median for the simulations to the observed distribution of
abundances about the median.  We chose a reference temperature of 6000K and
an [Fe/H] of $-2.3$ for the simulations and used the Thorburn (1994) mean trend
to correct the data to a uniform T$_{\rm eff}$ and [Fe/H].  Histogram plots
of the number of stars as a function of lithium abundance are compared to the
number of stars as a function of lithium depletion factor expected in the
simulations without and with observational errors in Figures 9 and 10
respectively.  The s0 case is a good match to the width of the distribution,
while the s0.3 case is in marginal agreement and the s1 case clearly has a
dispersion too large to be compatible with the data.  For comparison
purposes, an analogous distribution of depletion factors for solar abundance
models at a temperature of 6000K is also shown.

The dispersion in the bulk of the halo population is therefore consistent
with overall depletions in the range of 0.2-0.4 dex, with lower depletion
factors being favored.  However, the existence of highly overdepleted stars
can also be used to set bounds on the absolute depletion, and this favors
higher absolute depletion factors in the above range.  For the s0 case, the 
median
depletion is 0.2 dex and the maximum is 0.6 dex; stars more than 0.4 dex below
the plateau should therefore not be seen, in contradiction with the data.
The corresponding bounds for the s0.3 case are 0.4 and 1.2 dex, permitting
stars as much as 0.8 dex below the plateau; this is in better agreement with
the data.  We also note that the observed fraction of such stars (4 out of
90) is similar to the fraction of very rapid rotators in young open clusters.
We therefore conclude that the distribution of lithium abundances in the
plateau is consistent with a range of 0.2-0.4 dex absolute depletion, with
the bulk dispersion favoring lower values and the overdepleted stars favoring
higher values within this range.

\subsection{\6li Depletion}

In Figure 11 we show the relationship between \7li destruction 
and \6li destruction in the rotationally mixed models chosen to 
correspond to the hot halo subgiant HD 84937 ([Fe/H]$\approx  -2.2$  and 
\6li/\7li $\approx 0.06$) (\cite{SLN}; \cite{HT94},\cite{HT97}), to date the 
only star in which the detection of \6li has been confirmed by more than
one group (see Hobbs \& Thorburn (1997) for a detailed observational summary 
of the \6li data).   Note that, depending on the solar normalization, 
the ratio of \6li destruction to \7li destruction can vary from a few 
(for the s0 case) to 10 (for the s1 case).  The absolute \6li depletion factor 
varies between roughly 6 and 70, respectively.  Note that this is not the 
usual folklore of lithium 
depletion which anticipates substantial \6li destruction in any star where 
even minimal \7li destruction has occurred (\cite{BS88},\cite{VC98}).  Indeed,
 it has often been argued that 
because \6li has been observed in at least one halo star, there can be 
little, if any, \7li depletion in any \popii plateau stars (\cite{SFOSW}).  
However, this argument was based on standard models without rotational mixing 
where all the lithium destruction is by convective burning during the 
star's approach to the main sequence.  Then, since \6li is much more 
fragile than \7li, any star in which \6li is not significantly destroyed is 
not expected to destroy any \7li at all.  This \6li - \7li connection is 
relaxed in models which include rotational mixing.  
In the convective 
burning only models, all the material on the surface has been exposed 
to the same temperature whereas in models with rotational mixing 
different 
fractions of the surface material are exposed to different temperatures.  
Although the rank ordering (with binding energy) of the magnitude of 
lithium destruction is preserved (\ie, \6li is always destroyed more 
than \7li), the relative amounts of depletion can vary since the surface 
of the star is a mixture of gas which has been exposed to {\it different}
temperatures.  Note also that rotational models 
predict a dispersion of \6li depletion at fixed T$_{\rm eff}$ corresponding 
to different initial rotational velocities as drawn from a distribution 
of initial conditions.  Below we use the \popii \6li observations to 
further constrain models of rotational mixing and overall lithium 
depletion.

To use the observed abundance of \6li as a constraint 
on the amount of lithium depletion, we require an estimate of its initial 
abundance.  \6li is only produced in interesting 
(\ie, observable) amounts by cosmic ray nucleosynthesis via 
two channels: (1) cosmic ray proton and alpha spallation of CNO nuclei 
in the ambient ISM (and {\it vice-versa}) and, (2) alpha-alpha fusion 
reactions.    To avoid introducing more adjustable parameters 
than can be constrained by the observational data we can tie the CNO 
contribution of \6li to the observed abundances of B and Be which are 
produced by the same spallation process.  The Be and B abundances in 
\popii halo stars are observed to increase linearly with Fe/H 
(\cite{Dun97}) and, therefore, so too should the ``CNO-tagged'' \6li.  
The predicted ratio of spallation-produced \6li to Be or B depends on
the specific cosmic ray spectrum and relative target/projectile abundances 
adopted.  Some of these details can be avoided if we simply normalize \6li 
production to the solar abundance of \6li ([\6li]$_\odot \sim 2.2$): 
[\6li]$_{\rm CR(CNO)} = $ [\6li]$_{\odot}$ + [Fe/H] = 2.2 + [Fe/H].  This 
represents an upper limit to the CNO CR production of \6li at any epoch.  
Since the metallicity of the Spite plateau stars is [Fe/H] $\leq -1.3$, 
the CNO contribution will, in general, be quite small: [\6li]$_{\rm CR(CNO)} 
\leq 0.9$.  If this were the only production mechanism for \6li in halo stars, 
it would, along with the plateau value of [Li] $\sim$ 2.2, correspond to a 
ratio \6li/\7li $\le 1/20$, which is comparable to that observed.  In this 
case (production via spallation of CNO) there would be room for little, if 
any, \6li destruction.  

To quantify this point, consider the case of 
HD 84937, a hot ($T_{\rm eff} \sim 6300$K) metal-poor star in which \6li 
observations have been claimed by two independent groups
(\cite{SLN}; \cite{HT94}, 1997).  
They find [Fe/H] $= -2.2\pm 0.2 $ and [\6li]$_{\rm OBS} = 1.0 
^{+0.1}_{-0.2}$, an order of magnitude {\it more} \6li than the maximum predicted
from CNO spallation alone: [\6li]$_{\rm CR(CNO)} = 0.0\pm 0.2$.   
It must be emphasized that 
this can NOT be used as evidence for the absence of \6li destruction in 
halo stars but instead points to the necessity of an additional source for 
the observed \6li (in particular, one which does not scale with metallicity). 

Alpha-alpha fusion synthesis of \6li (\cite{SW}; SW),  is not necessarily 
coupled to the cosmic ray spallation production of Be and B and could, in 
principle, be such a source of \6li (and a comparable amount of \7li).  
Naively we expect the fusion contribution to scale as one lower power of 
Fe/H than does the CNO-spallation contribution (\ie, that it may be nearly
independent of metallicity).  As before for spallation, the normalization 
of the alpha-alpha \6li contribution is essentially arbitrary.  As an extreme 
upper limit to the abundance of alpha-alpha produced \6li in the gas out 
of which the \popii stars form, we may adopt the solar abundance:

\begin{equation}
[^6Li]_{CR(\alpha\alpha)} \le 2.2.
\end{equation} 
This bound is certainly robust (and most likely overly conservative) since it
assumes that CR fusion in the very early Galaxy generated \6li at the solar
abundance. Nevertheless, it represents the maximum amount of \6li any Pop II
star could contain.
Defining the \6li depletion factor, $D_6$, as: (\6li/H)$_{OBS} 
\equiv$ (\6li/H)$_{CR(\alpha\alpha)}/D_6$, we have for HD 84937

\begin{equation}
\log D_6 < 1.2^{+0.2}_{-0.1}.
\end{equation}
That is, there is a model indenpendent upper bound to \6li destruction of a 
factor of 14-25 for HD 84937.  The rotational 
models for HD 84937 which violate this constraint (\ie, destroy too much 
\6li) are the same ones which generate too much dispersion in the plateau 
abundance of \7li and do not correctly predict the behavior of the lithium 
abundances in open clusters.  This $D_6$ constraint allows us to obtain an 
independent upper bound on the depletion of \7li in \popii plateau stars 
(see Figure 11) of $0.5-0.6$ dex.  We comment here that although 
Lemoine \etal (1997) claim 
to have constrained $D_6$ to be strictly less than four, they did not 
include the fusion production of \6li via alpha-alpha which we have seen 
may be required if \6li has been detected in HD 84937.

\subsection{Relative Depletion of \6li and \7li}
Since post-BBN production of lithium can compete with surface depletion, 
lithium production and destruction are coupled.  Thus the dispersion in 
the \popii lithium data itself can be used to bound any contributions 
to the observed \popii lithium abundance from post-big bang production 
and/or stellar destruction.   Every 
\popii star should start with the same amount of BBN-produced \7li 
 and variations in post-BBN production and stellar destruction
will contribute to dispersion in the observed ``plateau" abundances.
Any observational limits to the dispersion provide a limit to the 
combination of CR-production and stellar depletion of lithium in the halo 
stars.  That is, for any halo star, the observed total (\7li {\it plus} \6li) 
lithium abundance can be expressed as 
\begin{equation}
 y_{Li}^{\rm OBS} = \frac{y_{7{\rm BBN}}+y_{7{\rm CR}}}{D_7} + \frac{y_{6{\rm CR}}}{D_6},
 \end{equation} 
where the various $y_7$'s represent the observed (OBS), {\rm BBN}, and cosmic 
ray produced (CR) number fractions of \7li relative to hydrogen.
The CR contribution to the observed lithium, with account taken 
of stellar depletion, may be bounded by the dispersion in the observed 
lithium abundances:
\begin{equation}
 \frac{y_{7{\rm CR}}}{D_7} + \frac{y_{6{\rm CR}}}{D_6} \le \Delta y_{Li}^{\rm OBS} \approx 
 0.8\times 10^{-10},
\end{equation}
where we have assumed that the {\rm BBN} contribution is fixed by the minimum 
of the plateau:
\begin{equation}
\frac{y_{7{\rm BBN}}}{D_7} = (y_{Li}^{\rm OBS})_{min} \approx  1.4\times10^{-10}.
\end{equation}
To bound $D_6/D_7$ we may then write for the (inverse of the) \6li fraction 
\begin{equation}
\left(\frac{Li}{^6Li}\right)_{\rm OBS} =
\frac{D_6}{D_7}\left(\frac{y_7}{y_6}\right)_{CR}\left[\frac{y_{7{\rm BBN}}}{y_{
7{\rm CR}}}
+1\right] +1.
\end{equation}
These relations can be used along with the observational data to bound 
the combinations $y_{7{\rm BBN}}/D_7$ and $y_{7{\rm CR}}/D_7$, respectively.  The 
6-to-7 ratio expected (SW) from CR nucleosynthesis is $(y_6/y_7)_{CR} 
\le 0.9$, 
leading to a bound on $D_6$/$D_7$: 
\begin{equation}
\frac{D_6}{D_7} \la 8.
\end{equation}
That is, the small dispersion of the \popii lithium data along with the 
\6li abundance limits the {\it relative} 6-to-7 depletion.  We caution that
this calculation relies on the measured \6li in only one star, HD 84937.  
We find that the same models with rotation which, earlier, failed to 
reproduce the open 
cluster lithium abundance patterns, that predicted too much 
dispersion in the \popii lithium plateau, and too much absolute
destruction of \6li, also violate this constraint on the relative 
depletions of \6li and \7li.  From all these -- independent -- 
constraints a consistent picture of non-zero but limited lithium 
depletion in the warm, metal-poor \popii plateau stars emerges.

In conclusion, we find that those models with rotational mixing which are
constrained to fit the lithium abundance patterns in open clusters 
require that the Sun be an anomalously fast early rotator which 
has overdepleted lithium relative to most stars of its mass and age.
Extending these same models to halo stars yields depletion patterns which
can be constrained by the lithium abundances observed in these stars.  
The predicted dispersion in lithium abundances and the predicted \6li 
depletion factors are consistent with the \popii data for the {\it 
same} solar normalization required by the open cluster data.  Thus, 
models which satisfy all of the above criteria suggest that halo stars 
{\it must} deplete their initial lithium by at least 0.2 dex, but by no more 
than 0.4 dex. The conclusion that halo stars must deplete lithium in 
this range relies only on the assumption that halo stars have the 
same distribution of initial rotational velocities as do the open 
cluster stars.  If halo stars do not rotate then standard models 
 would predict no lithium depletion except at low T$_{\rm eff}$, contrary to 
the data.  Alternatively,
significantly more lithium might be destroyed without increasing the
dispersion in the observed lithium abundances if every halo star had 
nearly identical angular momentum histories. 

Armed with our estimate of the lithium depletion factor relevant for
the Spite plateau halo stars we may use the data to bound the 
primordial abundance of lithium.
\begin{equation}
[Li]_{\rm P} = [Li]_{\rm OBS} + \log D_7
\end{equation}
Adopting for the Spite plateau abundance [Li]$_{\rm OBS} = 2.25 \pm 0.10$ 
 and for the depletion factor $\log$ D$_7 = 0.30 \pm 0.10$,  the
primordial abundance should lie within the full range 2.35 $\leq 
[Li]_{\rm P} \leq$ 2.75 or,
\begin{equation}
2.2\leq 10^{10}(Li/H)_{\rm P} \leq 5.6.
\end{equation}
Here we have assumed the errors in
both the observed lithium abundance and the depletion factor are uniformly
distributed with a width of 0.2 dex.  This results in a triangular likelihood
distribution for [Li]$_P$ which has a 95\% C.L. range for the
log of the primordial lithium abundance only 5\% smaller than the 
full range values we quote here.  Since the shape of the distributions we 
assume are, at best, 
an approximation, we quote the full range in [Li]$_p$ resulting from the 
convolution of the two distributions.

\section{Implications for BBN}

The predictions of standard BBN are uniquely determined by one parameter, 
the density of baryons, parameterized by $\eta$ - the baryon-to-photon 
ratio: $\eta_{10} = 273\Omega_{\rm B}h^{2}$ ($\Omega_{\rm B}$ is the 
ratio of the baryon density to the critical density and the Hubble 
parameter 
is H$_{0}$ = 100h km/s/Mpc; $\eta_{10} = 10^{10}\eta$).  For standard BBN 
the early universe is assumed to have been homogeneous and expanding 
isotropically and the energy density at the time nucleosythesis begins (about 
1 second after the big bang) is described by the standard model of particle 
physics ($\rho_{\rm tot} =\rho_\gamma + \rho_e + {\rm N}_\nu \rho_\nu$, 
where $\rho_\gamma$, $\rho_e$, and $\rho_\nu$ are the energy density of 
photons, electrons and positrons, and massless neutrinos (one species), 
respectively, and $N_\nu$ is the equivalent number of massless neutrino 
species which, in standard BBN is exactly 3).  In Figure 12 the primordial 
abundances predicted by standard BBN are shown as a function of $\eta$.  
The width of each curve reflects the $\pm 2\sigma$ uncertainty in the 
predictions.  

Our analysis of lithium depletion
by rotational mixing, when combined with the \popii lithium data leads to a
primordial lithium abundance in the range:
$2.2\leq 10^{10}(Li/H)_{\rm P} \leq 5.6$.
Comparing with the predictions of standard BBN (see Figure 12), lithium
identifies two possible ranges for $\eta$.  A ``low-$\eta$" branch: 
$0.8\leq\eta_{10} \leq$ 1.7 and a ``high-$\eta$" branch: 3.7 $\leq \eta_{10} 
\leq$ 9.0.  Once $\eta$ is identified the primordial abundances of 
deuterium and helium-4 (also, helium-3) follow from the predictions 
of standard BBN.

\subsection{Low-$\eta$}

The baryon density corresponding to the low-$\eta$ range is quite 
small: $0.003\leq\Omega_{\rm B}h^{2} \leq$ 0.006.  If the x-ray emission
from rich clusters of galaxies provides a ``fair" sample of the
universal fraction of baryons (\cite{White93}; \cite{SF95}; \cite{E97};
\cite{SHF97}) this argues strongly for a very
low density, open Universe: $\Omega = \Omega_{\rm B}/f_{\rm B} \la 
0.1h^{-1/2}$.  Such a low baryon density may also be in conflict with the
baryon density required to reproduce the Lyman$-\alpha$ forest data 
in large-scale structure simulations (Weinberg \etal 1997).  
This low-$\eta$ range predicts a high primordial 
deuterium abundance ($10^{5}(D/H)_{\rm P} \geq$ 17) which is inconsistent 
with the low deuterium abundance inferred for some high-redshift, 
low metallicity QSO absorption systems (\cite{BT96}; 
\cite{TFB96}; \cite{TBK96}; \cite{BT97}) but not with the high 
abundances inferred for some other systems (\cite{Songaila94}; 
\cite{Carswell94}; \cite{Hogan95};
\cite{Hogan96}; \cite{Songaila97}; \cite{Webb97}).  This low-$\eta$ 
range 
corresponds to a BBN-predicted helium abundance (Y$_{\rm P} \leq$ 0.235) 
which is in excellent agreement with that inferred from emission-line 
observations of the low-metallicity, extragalactic \hii regions (\cite{OS}; 
\cite{OSS96}; \cite{HOS97}).

\subsection{High-$\eta$}

Avoiding the minimum in the BBN lithium ``valley'' (see Figure 12), this
alternate branch favors a higher baryon density: 0.014 $\leq 
\Omega_{\rm B}h^{2} \leq$ 0.033 which, when combined with the x-ray
cluster baryon fraction estimate, permits a higher total density Universe:
$0.2 \la \Omega h^{1/2} \la 0.6$.  A baryon density in this range is
consistent with the Lyman$-\alpha$ forest data (Weinberg \etal 1997).  On 
this branch the primordial
deuterium abundance is predicted to be smaller (6.6 $\geq 
10^{5}(D/H)_{\rm P} \geq$ 1.1), consistent with some recent estimates from
high-z QSO absorbers.  However, the primordial helium abundance does 
pose a 
 challenge to 
standard BBN since 0.243 $\leq$ Y$_{\rm P} \leq$ 0.253 is predicted 
while the \hii region data suggest 0.230 $\leq$ Y$_{\rm P} \leq$ 0.238 
(but see also \cite{Izzy94}, \cite{Izzy96} and \cite{TI97} who derive $0.244
\pm 0.002$ from an independent data set).

\section{Summary}

The lithium abundances of stars have interesting implications for cosmology 
and for the theory of stellar structure and evolution.  The conclusions of 
this paper impact both of these areas.  Standard stellar models implicitly 
assume that a variety of physical processes known to occur in real stars can 
be neglected; for many purposes this is a good approximation.  There is now 
extensive evidence in \popi stars that the predictions of standard models 
are not in agreement with all the data.  Furthermore, work from a variety 
of investigators on physical effects not included in the standard models has 
concluded that there are physically well-motivated processes which can affect 
the surface lithium abundances of stars.  Standard models and the previous 
generation of nonstandard models predicted initial \7li abundances that
differ by up to a factor of ten from the same set of current \popii data, 
with very different implications for cosmology.  

We believe that mild envelope mixing driven by rotation is the most 
promising candidate for explaining the {\it complete} observational picture.  
An improved treatment of angular momentum evolution in low mass stars
now allows us to generate stellar models with rotation which are consistent
with the rotation rates observed as a function of mass and age.  We have
inferred the distribution of initial angular momenta in young clusters
and computed the distribution of lithium depletion factors as a function
of mass, composition, and age.  The resulting lithium depletion pattern is
in good agreement with both \popi and \popii data, and the distribution of
theoretical depletion factors is consistent with the distribution of
abundances.  For \popii stars in particular, the observed slope of the 
[Li]-T$_{\rm eff}$ relationship, the observed dispersion in abundance at fixed 
T$_{\rm eff}$, the existence of a small population of overdepleted stars,
and the simultaneous detection of \6li and \7li in one halo star are all 
consistent with the predictions of the theoretical models.  The primary
uncertainty in the absolute depletion of \7li is the initial angular
momentum of the Sun, which is used to calibrate the mixing diffusion
coefficients.

We have used the observed properties mentioned above to
set bounds on \7li depletion for \popii stars.  Those
models which fit the \popi data best predict a small but 
real depletion of lithium in the warm (T$_{\rm eff} \geq 5800$K), 
metal-poor ([Fe/H] $\leq -1.3$) \popii stars: 0.2 $\leq$ logD$_{7} 
\leq$ 0.4.  These same models are consistent with the very small 
observed dispersion in abundances about the plateau value and with 
the survival of some \6li.  These depletion factors are significantly
less than in previous models with rotational mixing (PDD, Chaboyer
\& Demarque 1994), and this difference can be directly attributed to 
the adoption of an improved treatment of angular momentum evolution.
The uncertainties in the \7li depletion factor can be reduced by a
combination of new data and improved stellar evolution models.  A large
set of accurate lithium abundances in old open clusters would enable us 
to calibrate the diffusion coefficients for mixing without relying on the
solar calibration.  Microscopic diffusion and rotational mixing should be 
considered together, although the work of Chaboyer \& Demarque (1994) indicated
that the predictions of such models are similar to those which include
rotation alone.  The observed \7li depletion pattern is relatively insensitive
to the treatment of internal angular momentum transport, with a factor
of ten in the time scale corresponding to less than a 10\% change in the
logarithmic lithium depletion factor in the work of PDD.  However, the solar 
rotation curve does provide evidence for angular momentum transport
from mechanisms not considered in the current paper, such as magnetic fields 
and/or internal gravity waves.  Lithium depletion in models which include
these effects and hydrodynamic mechanisms should be explored.  Intermediate
metallicity stars could also provide a valuable test of the dependence of
lithium depletion on metal abundance. 

When our constraints on 
lithium-depletion are combined with the abundances inferred from
observations of the Spite plateau halo stars ([Li]$_{\rm OBS} = 2.25 
\pm$ 0.10) we find for the  range in the abundance of 
primordial lithium: 2.35 $\leq$ [Li]$_{\rm P} \leq$ 2.75 (2.2 $\leq 
10^{10}$(Li/H)$_{\rm P} \leq$ 5.6).  Although lithium is not the
ideal baryometer, a comparison with the predictions of standard BBN
identifies two options: low-$\eta$ and high-$\eta$.  The low-$\eta$
branch suggests a very low baryon density, open Universe which may be in
conflict with the baryon density inferred from observations of the
Lyman-alpha forest.  Although the helium abundance predicted
for this low-$\eta$ option agrees with that inferred from \hii region
data, the very high predicted deuterium
abundance may not be consistent with observations.  In contrast, the
high-$\eta$ branch corresponds to a baryon density consistent with
the Ly-$\alpha$ data and permits a higher density Universe.  In this
case the predicted primordial deuterium abundance is in excellent 
agreement with the low deuterium QSO data  but a considerably higher 
primordial helium abundance is predicted than is inferred from some of the 
observational data.

\acknowledgements

This work is supported by the Department of Energy Contract No.\
DE-AC02-76-ER01545 at The Ohio State University and NASA ATP Grant
68169492.  TPW thanks Sylvie 
Vauclair, Con Deliyannis, Corinne Charbonnel, Francois Spite, Rafael Rebolo, 
and Paulo Molaro for useful conversations.
  

%
%


\newpage

\figcaption[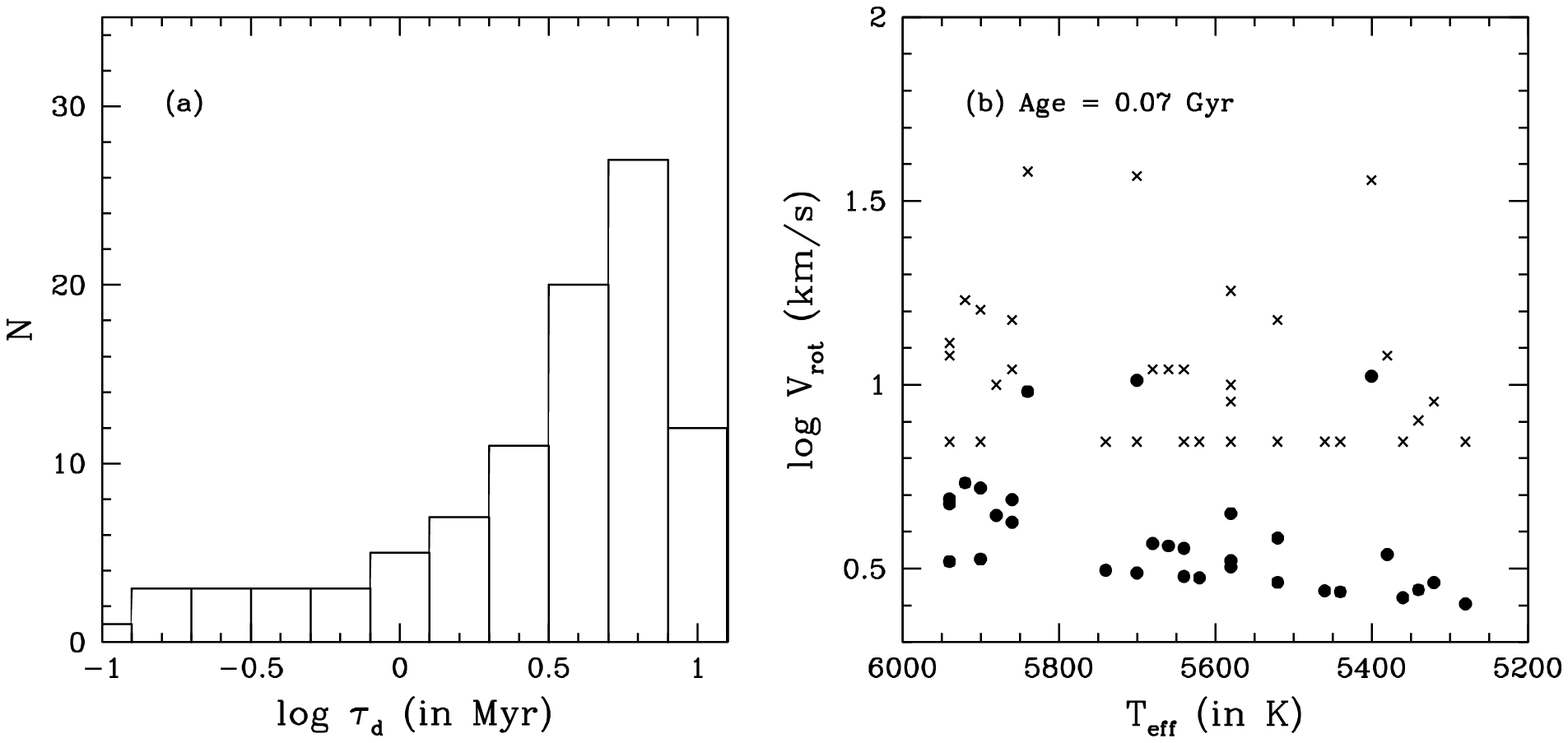]{
({\it a}) Distribution of accretion disk lifetimes inferred from the 
observed rotational velocities of Pleiades stars.  Data is taken from
Soderblom \etal (1993a).
({\it b}) Measured rotational velocities of the Pleiades stars (crosses)
in km/s
and the rotational velocities of the halo composition models (solid circles)
at an age of 70 Myr with the same $T_{\rm eff}$ and disk lifetimes.
}

\figcaption[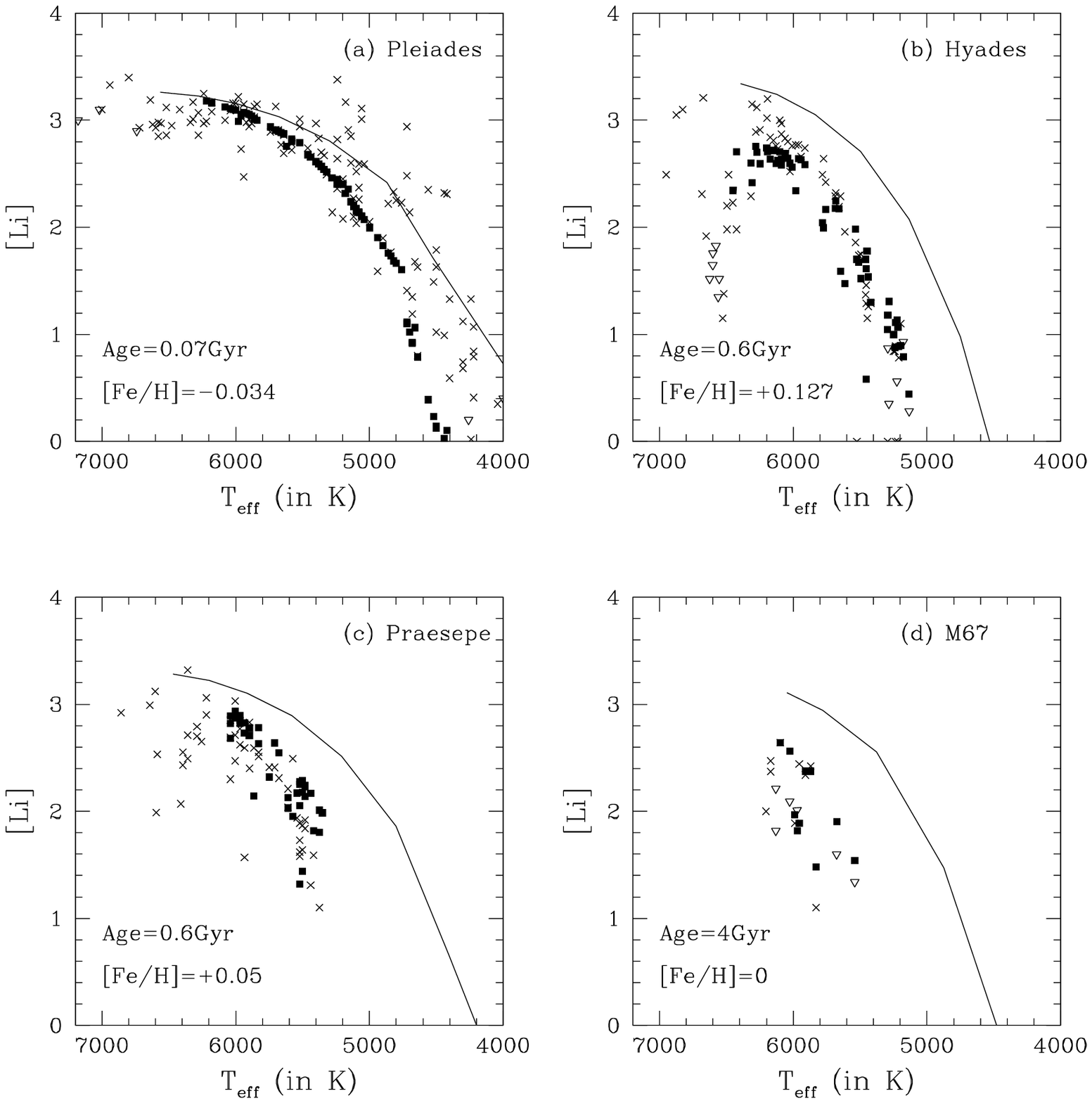]{
Comparison of the [Li] measured in four different open clusters (crosses)
 to the [Li] predicted by the s0 models (filled squares) in a typical
 Monte-Carlo run. The open inverted triangles represent those stars for
which only an upper limit is quoted in the observations.  Lithium data are
taken from Soderblom \etal (1993b) and Balachandran (1995).  The solid lines
are standard model \7li depletion factors for the metal abundances of the
clusters ([Fe/H]$=-0.03,+0.12,+0.04$, and 0.0 for the Pleiades, Hyades, 
Praesepe, and M67 respectively).  The initial abundance [Li] in the
theoretical models was
assumed to be 3.3 for M67 and 3.4 for the other clusters. 
}

\figcaption[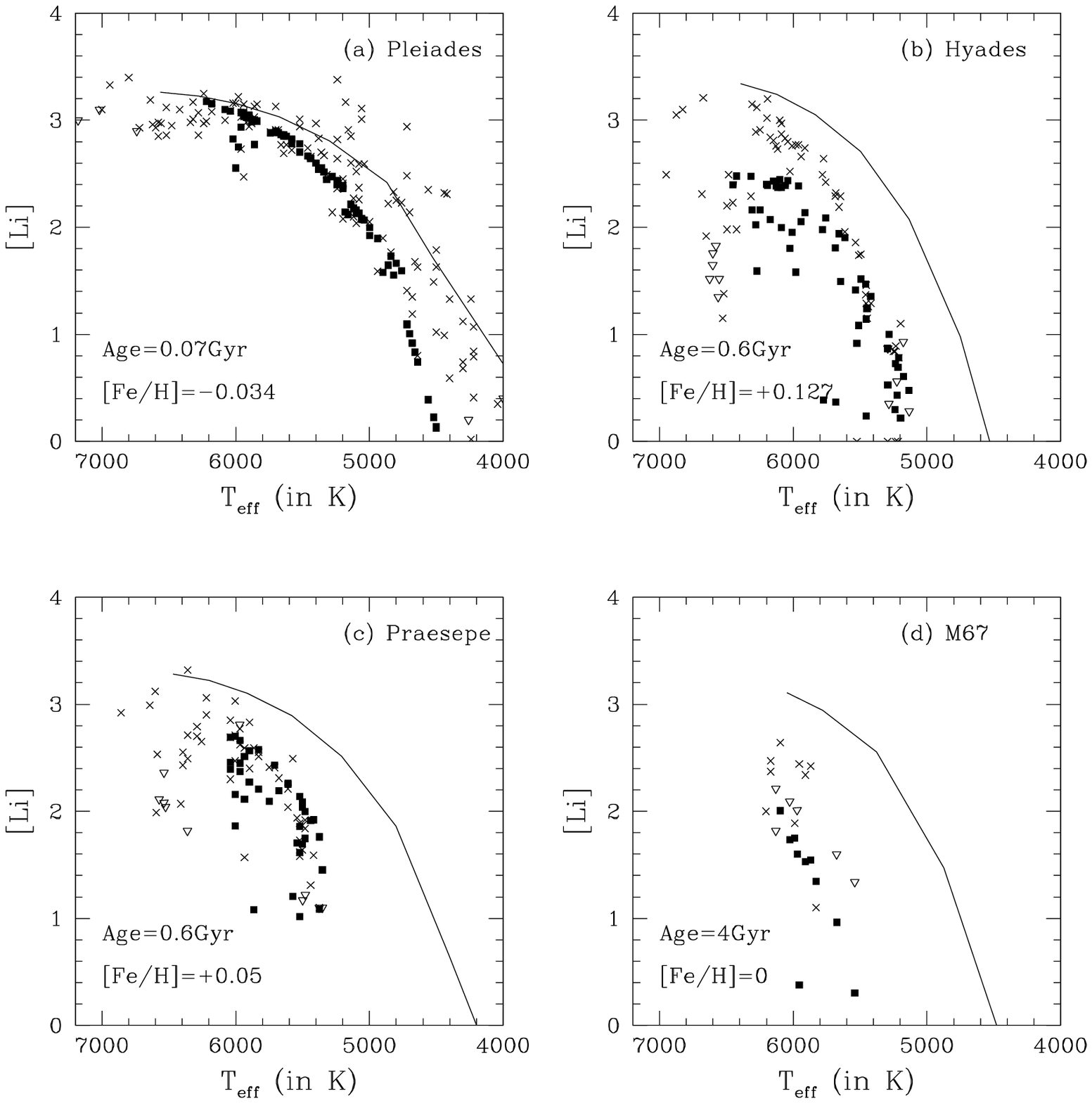]{
As Figure 2, except for a Monte Carlo simulation using the s1 case.
}

\figcaption[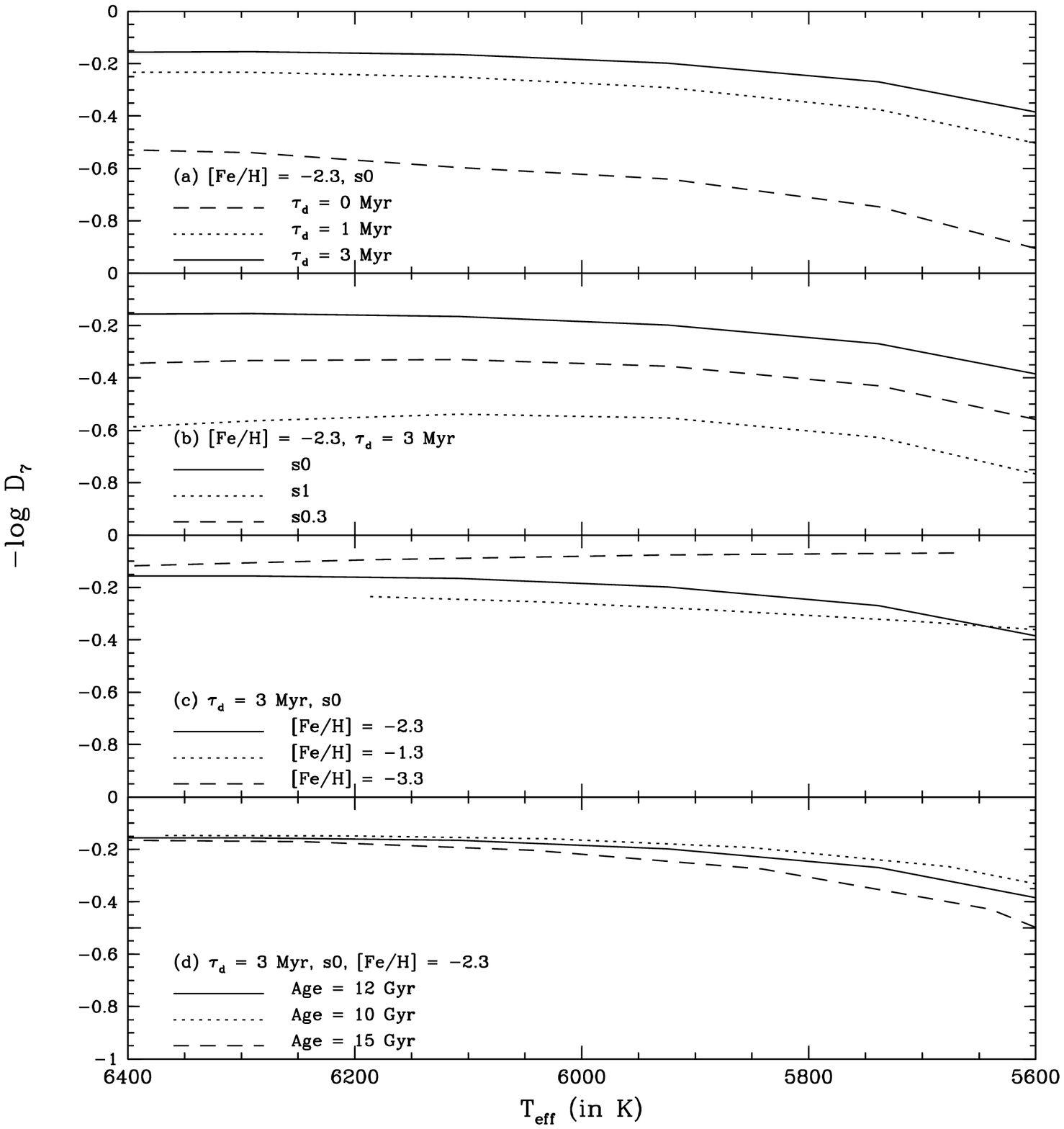]{
Sensitivity of the lithium depletion patterns to changes in 
({\it a}) the time scale for coupling between the accretion disk and
the protostar in Myr ($\tau_{d}$),
({\it b}) the solar calibration (accretion disk lifetime for the Sun),
({\it c}) [Fe/H] and 
({\it d}) age.
In all these panels, the solid line represents a base case for a typical 
halo star with $\tau_{d}$ = 3 Myr, [Fe/H] = $-2.3$, s0 and age = 12 Gyr.
$D_7$ is defined as the ratio of the current surface \7li to the initial
\7li.
}
\figcaption[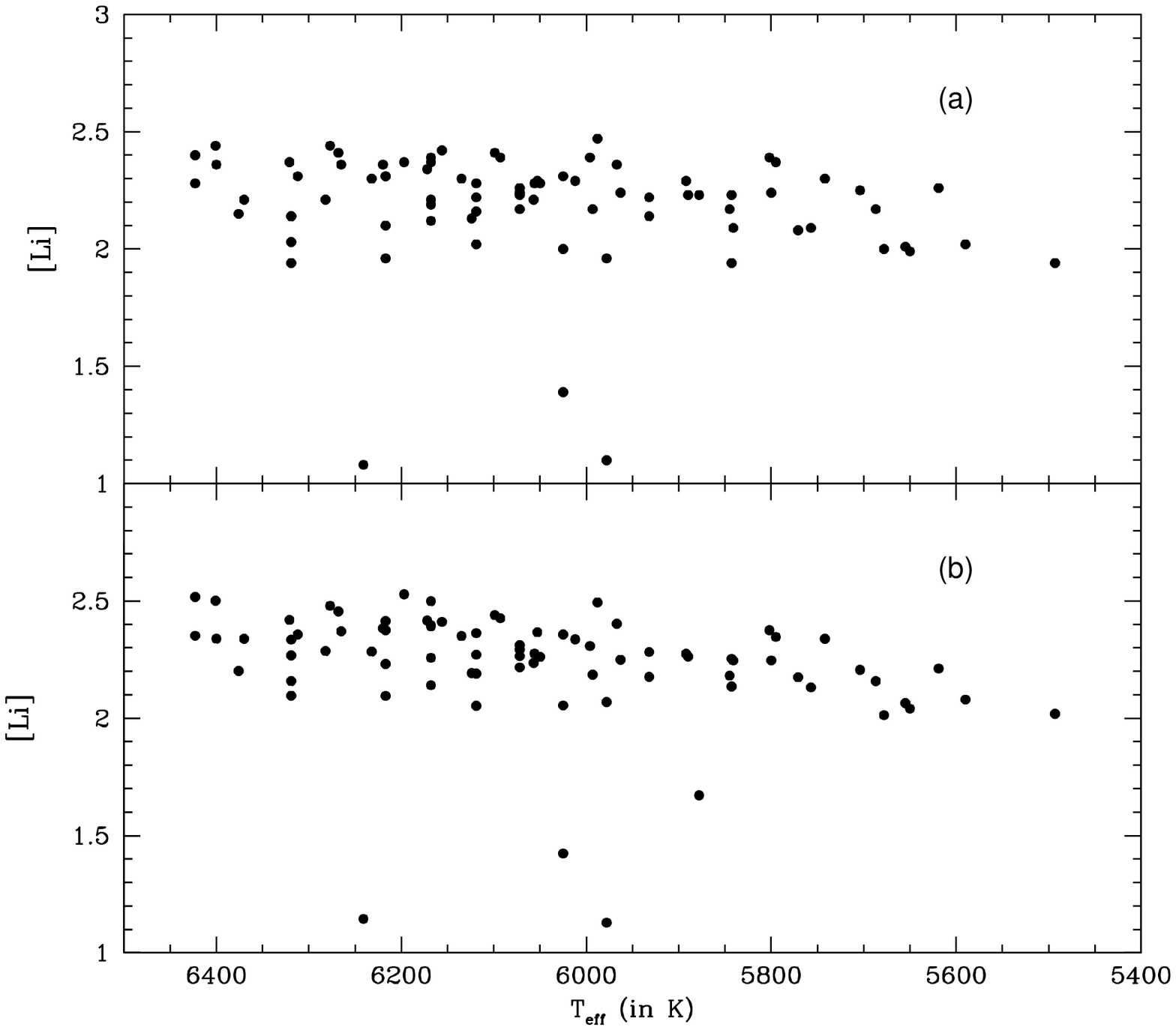]{
[Li] in halo stars before (panel a) and after (panel b) 
correcting for the metallicity trends.
The abundance data were taken from Thorburn (1994) and corrected to a
uniform metal abundance using equation 4.
}

\figcaption[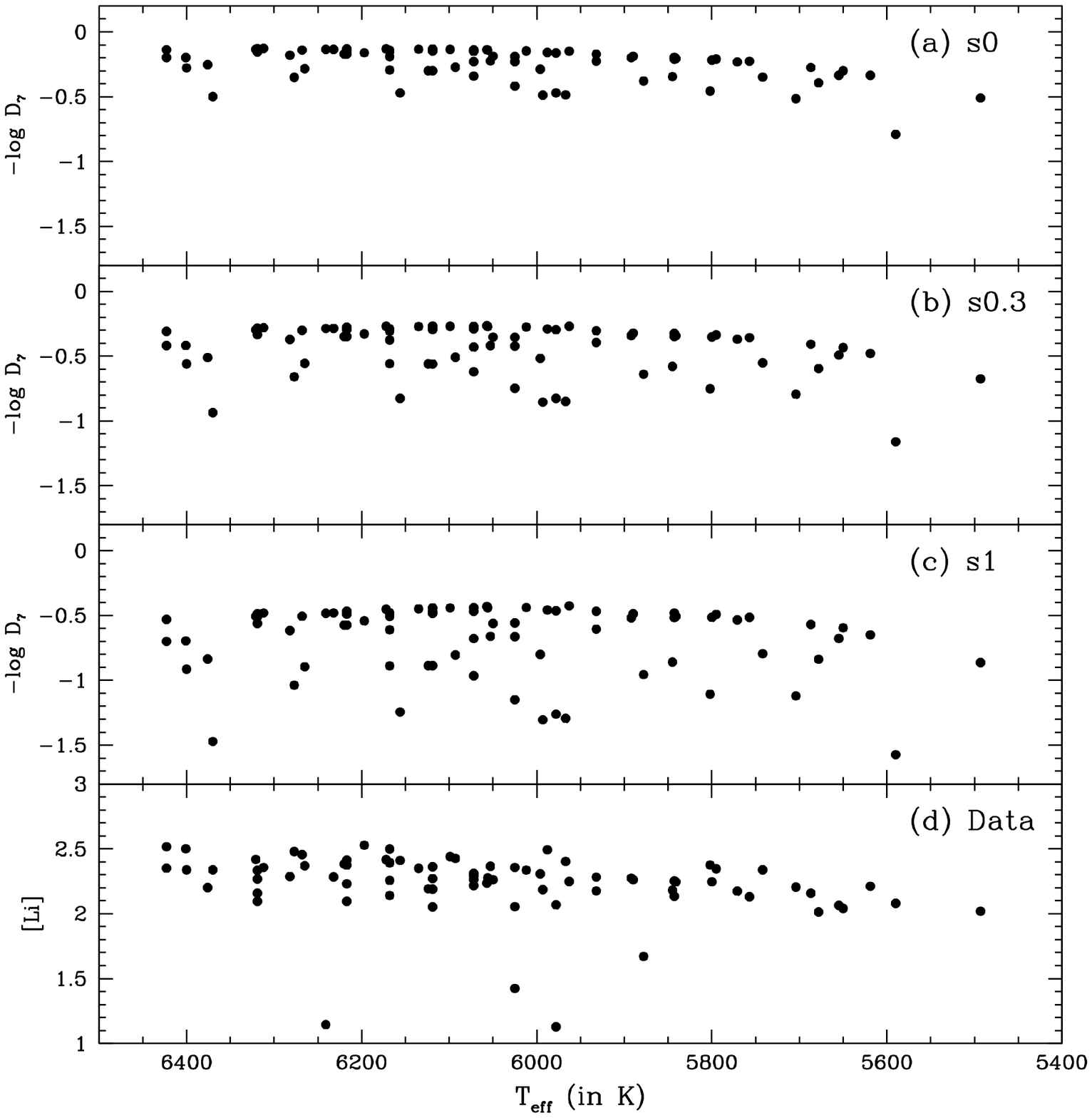]{
Sample Monte-Carlo distributions of \7li depletion factors
for the three different solar calibrations.
({\it a}) s0,
({\it b}) s0.3 and
({\it c}) s1.
({\it d}) Thorburn (1994) lithium data corrected to a uniform 
metallicity of [Fe/H] = $-2.3$.
}

\figcaption[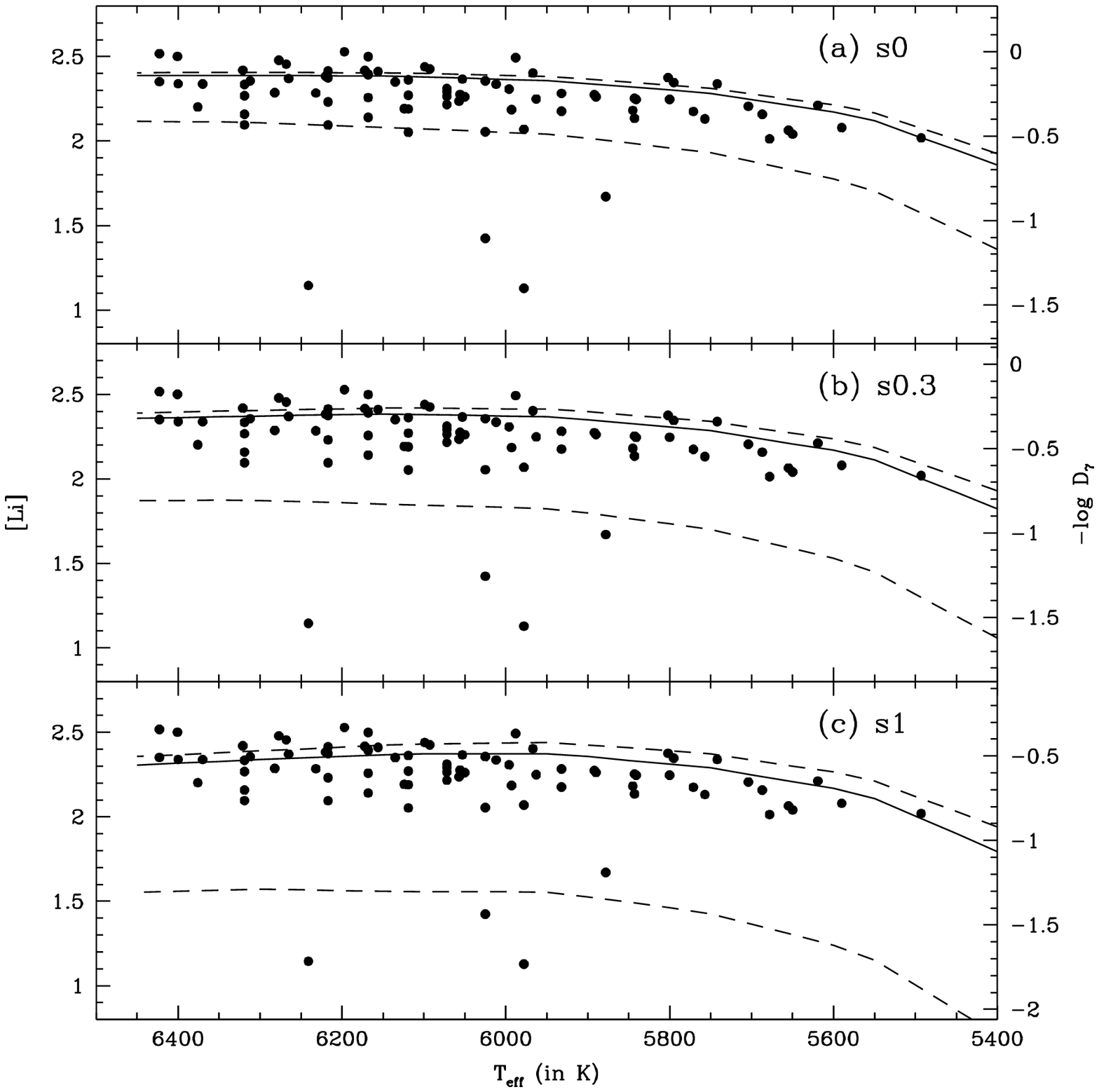]{
Lithium depletion factors for the three different 
solar calibrations before accounting for the observational errors.
({\it a}) s0,
({\it b}) s0.3 and
({\it c}) s1.
The solid line shows the median value, while the dashed lines show the 
2.5 and 97.5 percentile values.
The filled circles show the observed [Li] in the halo stars after 
correcting them all to the same metallicity of [Fe/H] = $-2.3$.
}

\figcaption[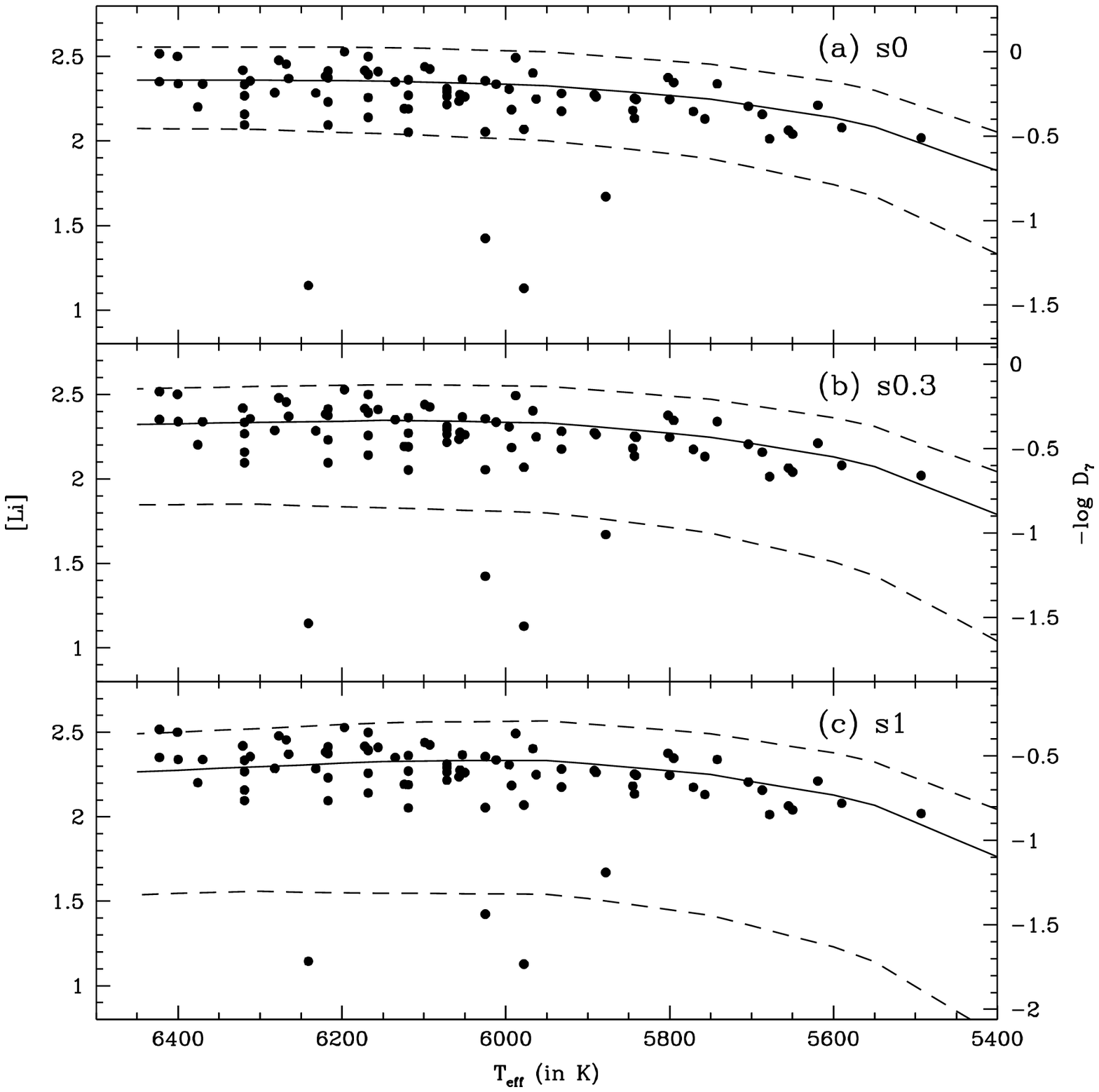]{
As Figure 7, except a random observational error of 0.09 dex was included
in the theoretical simulations.
}

\figcaption[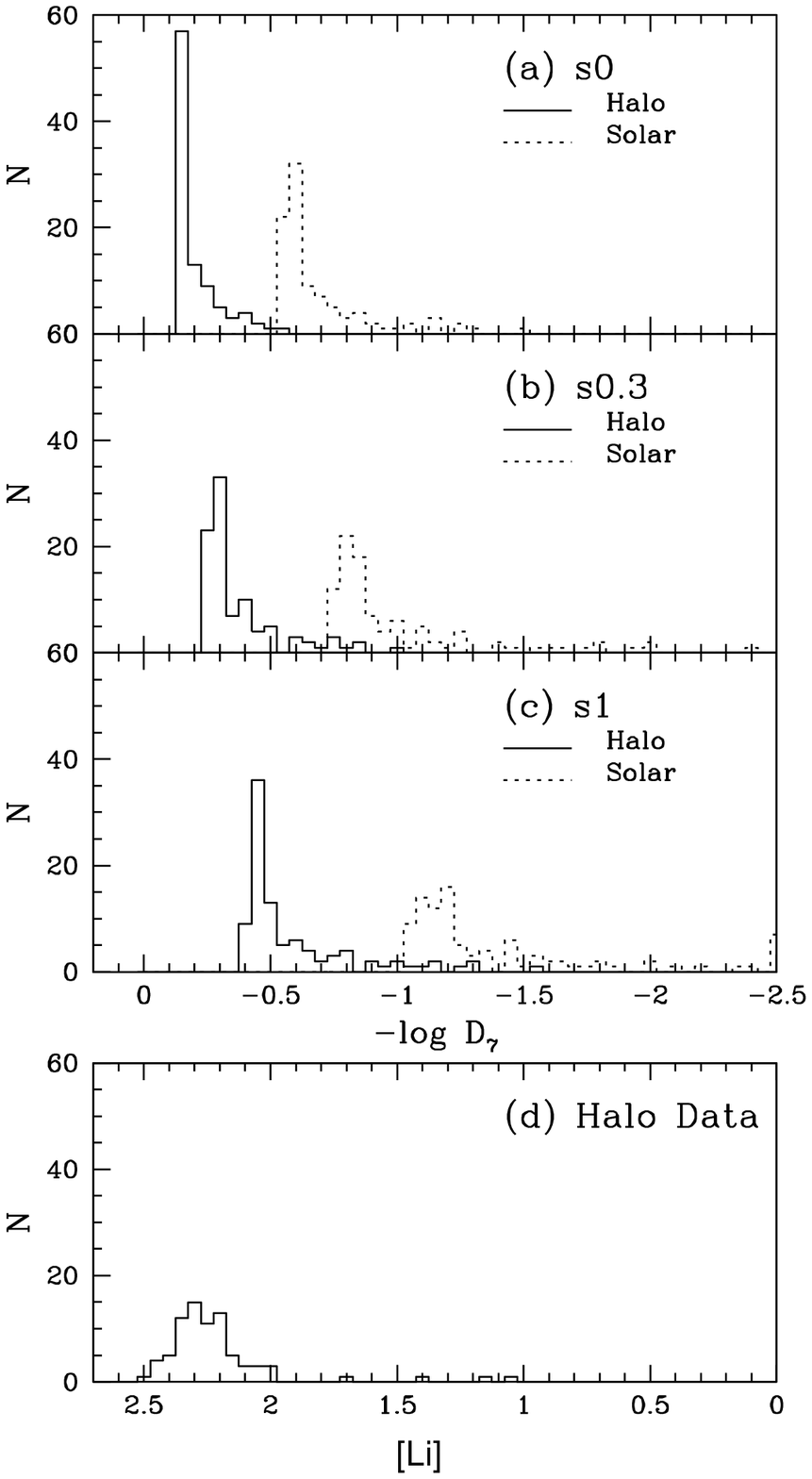]{
Distribution of the lithium depletion factors for the three different 
solar calibrations before accounting for the observational errors.
All the depletion factors are estimated at $T_{\rm eff}$=6000K and
[Fe/H]=$-2.3$. 
The top three panels show the results for the three different solar
calibrations.
({\it a}) s0,
({\it b}) s0.3 and
({\it c}) s1.
The dashed histograms are the distributions of depletion factors for the
same $T_{\rm eff}$ of 6000K but with a solar [Fe/H].  Panel ({\it d}) shows 
the observed distribution of [Li] in the halo stars after correcting them all 
to a metallicity of [Fe/H] = $-2.3$.  A correction of 
0.04($T_{\rm eff}-6000K$) was also applied to the data to correct for the
dependence of [Li] on effective temperature.
}

\figcaption[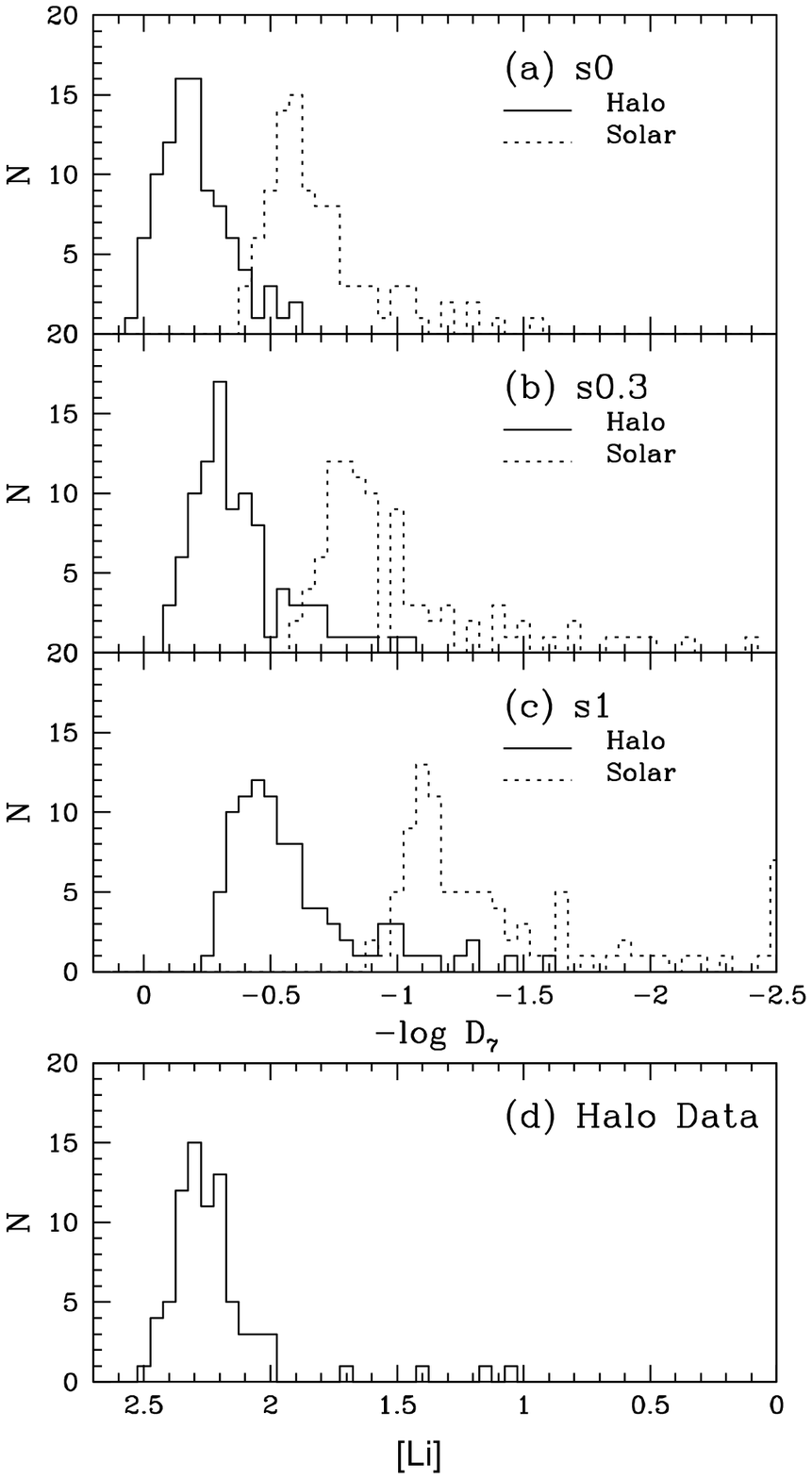]{
As Figure 9, except a random observational error of 0.09 dex was included in
the simulations.
}

\figcaption[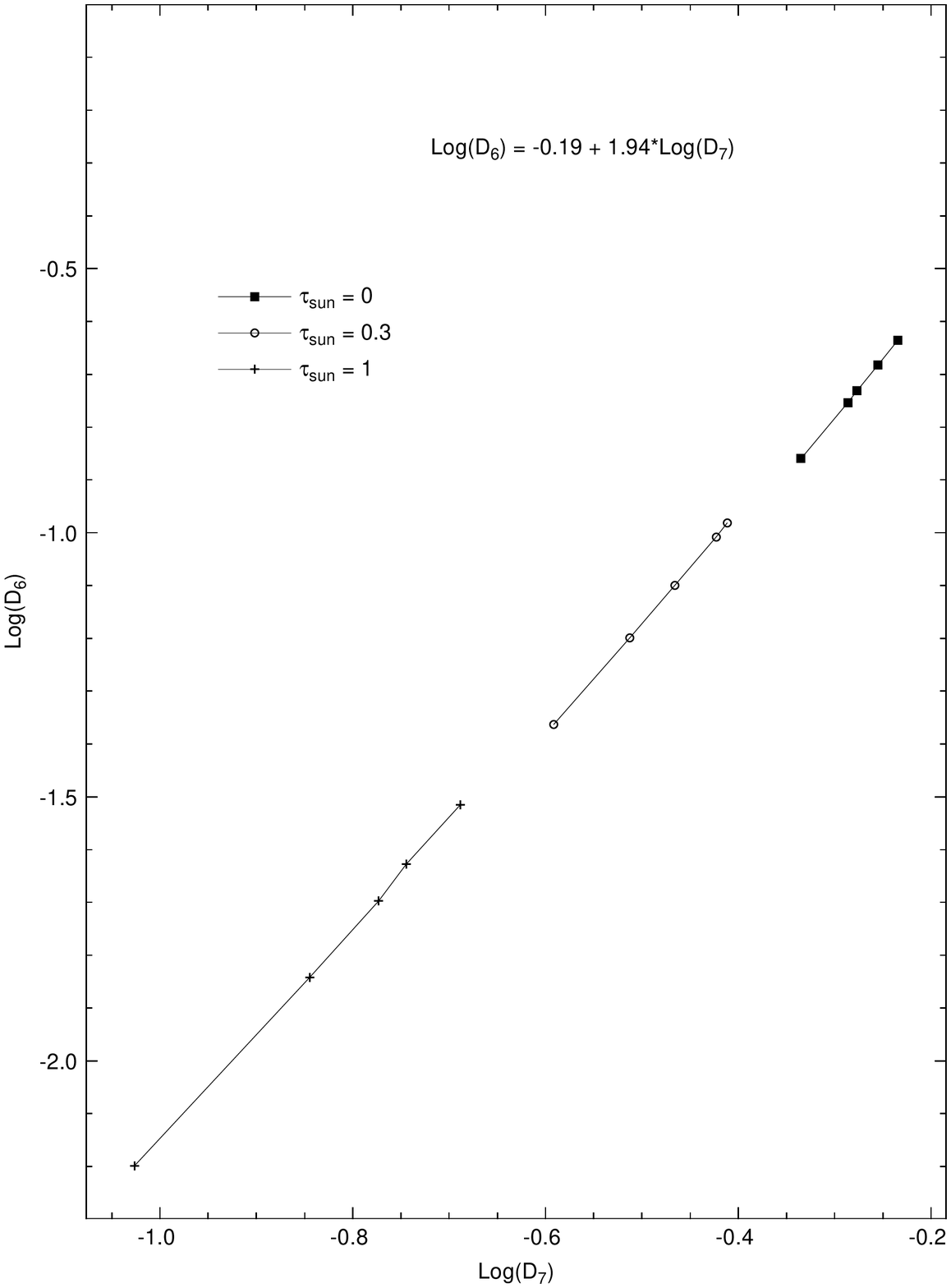]{
\6li depletion as a function of \7li depletion for models of HD 84937.  Three 
different solar calibrations are shown.  For each calibration the \6li and \7li 
depletion factors are indicated by the symbols (squares, open circles, crosses) 
for disk lifetimes of 1, 2, 3, 5, and 7 Myr (running from lower left to upper 
right respectively).  We have connected these discrete points by solid lines in
each case.  A least squares fit to \6li depletion as a function of \7li 
depletion for a disk lifetime of 3 Myr is also indicated. 
}

\figcaption[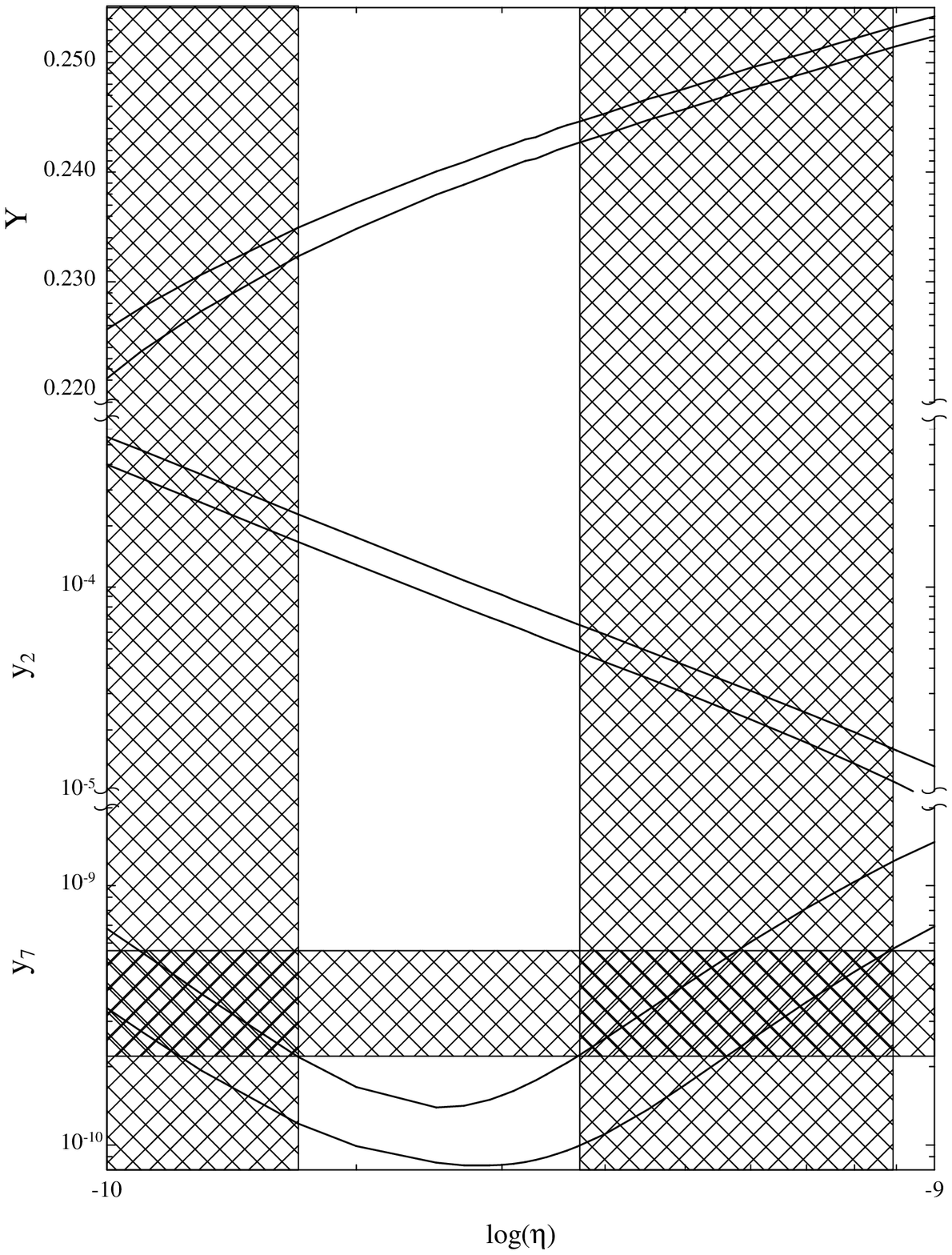]{
The predicted standard BBN yields (dashed lines represent 2-$\sigma$
uncertainties with 3 massless neutrino species) of light elements 
(\4he mass fraction, D and \7li number fraction relative to hydrogen) as a 
function of $\eta$, the baryon-to-photon ratio.  The primordial abundance 
of \7li, as inferred
from lithium abundances in Pop II stars and the analysis of stellar lithium 
depletion presented in this paper, is shown as a horizontal cross-hatched
band.  Vertical cross-hatched bands represent the allowed ranges of $\eta$
based on the primordial abundance of \7li.
}
 
\end{document}